\documentclass[aps,pra,twocolumn,groupedaddress,nofootinbib]{revtex4-2}
\usepackage{amsmath}
\usepackage{amsthm}
\usepackage{amssymb}
\usepackage{graphicx}
\usepackage{mathrsfs}
\usepackage{mathtools}
\usepackage{stmaryrd}
\usepackage{enumerate}
\usepackage{enumitem}
\usepackage[T1]{fontenc}
\usepackage[sc]{mathpazo}
\usepackage{physics}
\usepackage{subfigure}
\usepackage{overpic}
\usepackage{epstopdf}
\usepackage{bm}
\usepackage{color,soul}
\usepackage{tikz}
\usepackage[bookmarks=true]{hyperref}
\usepackage{xcolor}

\usepackage[linesnumbered]{algorithm2e}

\hypersetup{colorlinks=true,citecolor=blue,
linkcolor=blue,urlcolor=blue,pdfstartview=FitH,
bookmarksopen=true}
\newtheorem{theorem}{Theorem}

\usepackage{cleveref}
\crefname{equation}{Eq.}{Eqs.}
\Crefname{equation}{Equation}{Equations}
\crefname{theorem}{Theorem}{Theorems}
\crefname{lemma}{Lemma}{Lemmas}
\crefname{appendix}{Appendix}{Appendixes}
\crefname{figure}{Fig.}{Figs.}
\crefname{section}{Sec.}{Secs.}
\Crefname{section}{Section}{Sections}
\crefname{algorithm}{Algorithm}{Algorithms}

\begin{document}

\title{Retrieving maximum information of symmetric states from their corrupted copies}
\author{Zhao-Yi Zhou}
\affiliation{Department of Physics, Shandong University, Jinan 250100, China}
\author{Da-Jian Zhang}
\email{zdj@sdu.edu.cn}
\affiliation{Department of Physics, Shandong University, Jinan 250100, China}
\date{\today}

\begin{abstract}
    Using quantum measurements to extract information from states is a matter of routine in quantum science and technologies. A recent work [\href{https://doi.org/10.1103/PhysRevLett.133.040202}{Phys. Rev. Lett. 133, 040202 (2024)}] reported the finding that the symmetric structures of a state can be harnessed to dramatically reduce the sample complexity in extracting information from the state. However, due to the presence of noise, the actual state at hand is often corrupted, making its symmetric structures distorted before the execution of quantum measurements. Here, using the methodology of quantum metrology, we identify the optimal measurement that can retrieve maximum information of a symmetric state from its corrupted copies. We show that this measurement can be found by solving a semidefinite program in generic cases and can be explicitly determined for a large class of noise models covariant under the symmetry group in question. The results of this study nicely complement the recent work by providing a method to optimally utilize the distorted symmetric structures of corrupted states for information retrieval.
\end{abstract}

\maketitle

\section{Introduction}
Using quantum measurements to extract information from quantum states is an indispensable ingredient of quantum information processing, underpinning numerous applications across quantum science and technologies. One primary example is to extract the expectation value $\overline{X}:=\mathrm{tr}(\rho X)$ of an observable $X$ in a state $\rho$ from the quantum measurements performed on $\rho$. A long-standing pursuit in this line of research is to devise efficient methods to reduce sample complexity in quantum measurements, which has led to the proposals of compressed sensing \cite{GLF10,LZL12}, adaptive tomography \cite{Mahler2013PRL,Qi2017nQI}, self-guided tomography \cite{Ferrie2014PRL,Rambach2021PRL}, and classical shadows \cite{Aaronson2018,HKP20,HKP21}.

The recent work \cite{ZT24} explored leveraging symmetric structures of states to reduce the sample complexity in measuring expectation values of observables. The state $\rho$ is said to be symmetric under a group $G$ if it satisfies
\begin{equation}\label{symmetric-structure}
    U_g\rho U_g^\dagger=\rho ~~\textrm{for}~~ g\in G,
\end{equation}
where $U_g$ denotes a unitary representation of $G$. This equation
defines the symmetric structures of $\rho$, which are pervasive in quantum physics and frequently encountered in diverse contexts. A salient example showcasing the emergence of symmetric structures arises in condensed-matter physics, where the states of interest commonly exhibit translational symmetries \cite{Zhou2023E}. Another well-known example is in multipartite experiments \cite{KST07,WSK08,WKK09,PCT09,KWO10,EMKZ18,LZM21}, where the states under consideration often remain invariant under permutations \cite{TWG10,MHT12,GYvE14}. Notable instances of permutation-invariant states include the Werner states \cite{Wer89}, the Dicke states \cite{Dic54}, and the Greenberger-Horne-Zeilinger (GHZ) states \cite{GHZ89}, which are useful resources in quantum information processing tasks \cite{HHHH09,GT09,Hu2018PR}.

The main finding of Ref.~\cite{ZT24} is that, when the state $\rho$ in question exhibits some symmetric structures, the optimal measurement for obtaining $\overline{X}$ is the projective measurement of another observable $Y$ rather than $X$ itself. Here,
\begin{equation}\label{Y}
    Y=\mathcal{P} \left( X \right),
\end{equation}
where $\mathcal{P}$ is the so-called $G$-twirling operation, defined as $ \mathcal{P} \left( X \right) =\sum_g{U_gXU_{g}^{\dagger}}/\left| G \right|$ for a finite group $G$, with $\left| G \right|$ the cardinality of $G$. When $G$ is a compact Lie group,  $\mathcal{P} \left( X \right) =\int_G{d\mu \left( g \right) U_gXU_{g}^{\dagger}}$, where $\mu(g)$ is the normalized Haar measure \cite{Hal50, Mel24}. Two key properties of $Y$ are that
\begin{eqnarray}\label{property1}
    \overline{Y}=\overline{X},
\end{eqnarray}
but
\begin{eqnarray}\label{property2}
    (\Delta Y)^2\leq(\Delta X)^2,
\end{eqnarray}
where $(\Delta X)^2:=\tr(\rho X^2)-(\tr\rho X)^2$ is the quantum uncertainty of $X$ and $(\Delta Y)^2$ is defined in a similar way. Physically, the equality (\ref{property1}) means that the projective measurement of $Y$ can be an alternative to the projective measurement of $X$ for obtaining $\overline{X}$. The inequality (\ref{property2}) implies that the former generally consumes fewer samples than the latter for reaching the same measurement precision.

\begin{figure}
    \centering
    \includegraphics[width=\linewidth]{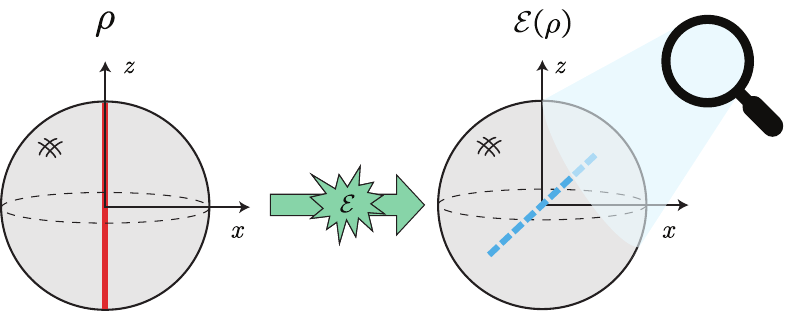}
    \caption{Schematic illustration of the setting under consideration. Consider the symmetric structures described by the group $G=\{\mathbb 1, \sigma_z\}$, where $\mathbb 1$ denotes the identity matrix and $\sigma_z$ denotes the Pauli-$Z$ matrix. A state $\rho$ exhibits these symmetric structures if and only if it can be parametrized as $\rho=\textrm{diag}(\theta, 1-\theta)$, which corresponds to a point located on the vertical red (solid) line in the Bloch sphere. The presence of noise corrupts the state $\rho$, transforming it to be $\mathcal{E} \left( \rho \right)$. For example, when $\mathcal{E}(\rho)=\rho /2+H\rho H/2$ with $H$ the Hadamard gate, $\mathcal{E}(\rho)$ is represented by a point located on the blue oblique (dashed) line in the Bloch sphere. The shift in the location indicates the distortion of the symmetric structures in question. Our purpose is to find the optimal measurement that is performed on $\mathcal{E}(\rho)$ and can take advantage of the distorted symmetric structures in $\mathcal{E}(\rho)$ to extract the information about $\rho$.}
    \label{fig1}
\end{figure}

The purpose of the present study is to complement the recent work \cite{ZT24} by taking into account noise, which can be, without loss of generality, modeled by a completely positive and trace-preserving map $\mathcal{E}$. As a persistent topic in quantum information science \cite{Pre18,JWBA23}, the state $\rho$ may be corrupted by noise before the execution of quantum measurements \cite{ZT22,UNM23}. This leads to the fact that the actual state available in the presence of noise is the corrupted state $\mathcal{E}(\rho)$ rather than $\rho$ itself (see Fig.~\ref{fig1} for a schematic illustration). A natural question then arises: How can we optimally measure the expectation value $\overline{X}=\tr(\rho X)$ of $X$ in $\rho$ when the actual state at hand is $\mathcal{E}(\rho)$? This question is highly nontrivial as the symmetric structures described by Eq.~(\ref{symmetric-structure}) are distorted in $\mathcal{E}(\rho)$.

In this study we answer this question. The information content we use is the quantum Fisher information (QFI) \cite{Braunstein1994PRL,Zhang2020PRR}, the inverse of which characterizes the optimal sample complexity in quantum measurements according to the celebrated quantum Cram\'{e}r-Rao bound \cite{ZT24}. Moreover, it makes sense to say that the QFI is the maximal information extractable via a quantum measurement, as the classical Fisher information naturally characterizes the information extracted from a measurement and the QFI, by definition, is the maximal classical Fisher information over all measurements. To identify the optimal measurement capable of extracting the QFI, we resort to the geometric formulation of parameter estimation theory \cite{TAD20}, which is well suited for dealing with the estimation of a parameter that can be expressed as a function of $\rho$, such as the expectation value of an observable, the fidelity to a given pure state, and the von Neumann entropy. Resorting to this theory, we show that the optimal measurement can be found by solving a semidefinite program (SDP) whenever the noise model $\mathcal{E}$ is invertible.\footnote{As a linear map on the space of Hermitian operators, $\mathcal{E}$ can be described by a matrix if we choose a basis for this space. By saying that $\mathcal{E}$ is invertible, we mean that the matrix associated with $\mathcal{E}$ is invertible. Intuitively speaking, a non-invertible $\mathcal{E}$ arises in the situation that the information about the initial state $\rho$ is lost and cannot be retrieved \cite{ZJZ24}. As the aim of this paper is to retrieve information from corrupted copies of $\rho$, we focus on the setting that $\mathcal{E}$ is invertible.} However, it turns out that this measurement may depend on $\rho$, due to which a refined knowledge of $\rho$ may be required in order to implement it in practice. We clarify that such an unpleasant dependence issue is a common feature in quantum metrology rather than being exclusive to the present study.

To release the requirement on the knowledge of $\rho$, we further specialize our discussions to a large class of noise models that are covariant under the symmetry group $G$. The studies on covariant quantum operations have been extensive and garnered significant interest because of their relevance in various physical contexts. For example, the absence of a quantum reference frame such as a phase or Cartesian reference frame imposes constraints on a party's ability to prepare states and perform quantum operations. This has sparked a line of development known as quantum reference frames \cite{BRS07}, where covariant quantum operations are those that can be executed without access to a reference frame. Another example is the presence of superselection rules \cite{WWW52,BW03}, which restricts the permissible quantum operations on a quantum system to be covariant ones. More generally, the presence of symmetries in a system generally imposes restrictions on the manipulation of the system, which results in nontrivial limitations on the implementation of quantum operations \cite{MS14,PCB16,ZYHT17}. This has led to the proposal of the resource theory of asymmetry \cite{GS08,CG19}, where covariant quantum operations are the free operations that do not consume or increase the asymmetry resource of states.
We show that the optimal measurement can be explicitly determined for covariant models, thereby eliminating the dependence issue mentioned above.

Finally, we apply our results to an experimentally relevant scenario, demonstrating how to optimally take advantage of the distorted symmetric structures in $\mathcal{E}(\rho)$ to reduce the sample complexity in information retrieval.

This paper is organized as follows. In \cref{sec_preliminaries}, we set the stage of our analysis. In \cref{sec_optstr}, we show how to find the optimal measurement whenever the noise model $\mathcal{E}$ is invertible. In \cref{sec_covchannel}, we specialize our discussion to the noise models that are covariant under the group $G$. We apply our results to an experimentally relevant scenario in \cref{sec_application} and conclude this paper in \cref{sec_conclude_discuss}.

\section{Stage of our analysis} \label{sec_preliminaries}

We are interested in the QFI about $\overline{X}$ given $\mathcal{E}(\rho)$ \cite{ZT24}, denoted as $\mathcal{J}[\overline{X}; \mathcal{E}(\rho)]$, which represents the maximal information about $\overline{X}$ that we can extract from $\mathcal{E}(\rho)$ via a quantum measurement \cite{Hel76,Hol11}. To specify the form of $\mathcal{J}[\overline{X}; \mathcal{E}(\rho)]$, we recall that the $\rho$ satisfying Eq.~(\ref{symmetric-structure}) can be parametrized using the representation theory of groups \cite{ZT24}. That is, $\rho$ can be expressed as $\rho(\bm{\theta})$ for some parameters $\bm{\theta}=(\theta_1,\cdots,\theta_q)$, where $q$ denotes an integer. For example, a qubit state respecting the symmetry group $G=\{\mathbb 1, \sigma_z\}$ can be expressed as $\rho=\textrm{diag}(\theta, 1-\theta)$, where $\sigma_\alpha$, $\alpha=x, y, z$, denote the Pauli matrices. The explicit form of $\rho(\bm\theta)$ in general can be found in Supplemental Material of Ref.~\cite{ZT24} and is presented in Appendix \ref{appsec_rho} of the present study, too.
Apparently, $\mathcal{E}(\rho)$ also depends on $\bm\theta$ and $\overline{X}=\tr(\rho X)$ can be regarded as a function of $\bm{\theta}$. We can express $\mathcal{J}[\overline{X}; \mathcal{E}(\rho)]$ as \cite{ZT24}
\begin{eqnarray}
    \mathcal{J}[\overline{X}; \mathcal{E}(\rho)]=1\bigg/\left(\partial\overline{X}^TH^{-1}\partial\overline{X}\right),
\end{eqnarray}
where $\partial\overline{X}:=(\frac{\partial\overline{X}}{\partial\theta_1}, \cdots, \frac{\partial\overline{X}}{\partial\theta_q})^T$ is a $q$-dimensional vector, and $H$ denotes the QFI matrix whose $ij$ element is given by
\begin{eqnarray}
    H_{ij}=\tr[\mathcal{E}(\rho) \frac{S_iS_j+S_jS_i}{2}].
\end{eqnarray}
Here $S_i$ is known as symmetric logarithmic derivative (SLD) \cite{Hel76,Hol11}, defined as the Hermitian operator that satisfies
\begin{eqnarray}\label[def]{SLD}
    \frac{\partial}{\partial\theta_i}\mathcal{E}(\rho)=\frac{1}{2}\left[\mathcal{E}(\rho)S_i+S_i\mathcal{E}(\rho)\right].
\end{eqnarray}
Below, we find a convenient formula for calculating $\mathcal{J}[\overline{X}; \mathcal{E}(\rho)]$. We do this by following the theory in Ref.~\cite{TAD20}.

Throughout this study, we only consider Hermitian operators unless otherwise specified.
We define a weighted inner product\footnote{Strictly speaking, this definition represents a pre-inner product as the positive-definiteness requirement may not be met when $\mathcal{E}(\rho)$ is singular.} between two operators $h_1$ and $h_2$ as
\begin{eqnarray}\label{inner-product}
    \langle h_1, h_2\rangle _{\mathcal{E}(\rho)}\coloneqq \mathrm{tr}\left[ \mathcal{E}(\rho) \frac{h_1h_2+h_2h_1}{2} \right],
\end{eqnarray}
where the subscript $\mathcal{E}(\rho)$ is used to indicate the dependence of this definition on $\mathcal{E}(\rho)$. Equation (\ref{inner-product}) induces a norm
\begin{eqnarray}
    \norm{h}_{\mathcal{E}(\rho)}=\sqrt{\langle h, h\rangle _{\mathcal{E}(\rho)}},
\end{eqnarray}
which inherits the dependence on $\mathcal{E}(\rho)$ from the inner product.

We introduce a linear space of zero-mean operators as
\begin{eqnarray}\label{def-Z}
    \mathcal{Z}_{\mathcal{E}(\rho)}=\{h: \tr[\mathcal{E}(\rho)h]=0\}.
\end{eqnarray}
It is easy to see that all the SLDs defined by Eq.~(\ref{SLD}) belong to $\mathcal{Z}_{\mathcal{E}(\rho)}$. Therefore, the linear span of these SLDs
\begin{equation}\label[def]{tangent-space}
    \mathcal{T} _{\mathcal{E}(\rho)}=\mathrm{span}_\mathbb{R}\left\{ S_1, S_2, \cdots, S_q \right\}
\end{equation}
is a subspace of $\mathcal{Z}_{\mathcal{E}(\rho)}$. $\mathcal{T} _{\mathcal{E}(\rho)}$ is known as the tangent space in the estimation theory in Ref.~\cite{TAD20}. We also introduce
\begin{eqnarray} \label{eq_def_t_Erhobot}
    \mathcal{T} _{\mathcal{E}(\rho)}^\bot=\{h\in \mathcal{Z}_{\mathcal{E}(\rho)}: \langle h, S_i\rangle_{\mathcal{E}(\rho)}=0, ~~i=1, \cdots, q  \},
\end{eqnarray}
which is the orthogonal complement of $\mathcal{T} _{\mathcal{E}(\rho)}$ in the space $\mathcal{Z}_{\mathcal{E}(\rho)}$.

\begin{figure}
    \centering
    \includegraphics[width=0.5\linewidth]{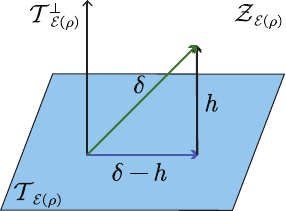}
    \caption{Schematic of the introduced concepts. The space $\mathcal{Z} _{\mathcal{E}(\rho)}$ is the direct sum of the tangent space $\mathcal{T}_{\mathcal{E}(\rho)}$ and its orthogonal complement $\mathcal{T}_{\mathcal{E}(\rho)}^\bot$. Accordingly, the influence operator $\delta$ can be decomposed as $\delta=(\delta-h)+h$, where $h$ is required to belong to $\mathcal{T}_{\mathcal{E}(\rho)}^\bot$. Such decompositions are not unique. The inverse of the quantum Fisher information $\mathcal{J}[\overline{X}; \mathcal{E}(\rho)]$ corresponds to the minimal length of $(\delta-h)$ over all such decompositions.}
    \label{fig_Hilbert_rho}
\end{figure}

A useful notion in the theory \cite{TAD20} is the so-called influence operator, defined as an operator $\delta$ in $\mathcal{Z}_{\mathcal{E}(\rho)}$ that satisfies
\begin{equation} \label{eq_delta_locallyunbiased_condition}
    \langle S_i,\delta \rangle_{\mathcal{E}(\rho)}=\frac{\partial}{\partial\theta_i}\overline{X},
\end{equation}
for $i=1,...,q$. Such an operator may not be unique. We choose
\begin{equation} \label{eq_construct_delta}
    \delta =\mathcal{E} ^{*-1}\left( X \right) -\overline{X} \mathbb{1},
\end{equation}
with $\mathbb{1}$ denoting the identity matrix.
Here we have assumed that $\mathcal{E}$ is invertible. $\mathcal{E}^*$ denotes the dual map\footnote{Let $\mathcal{E}(\rho)=\sum_i K_i\rho K_i^\dagger$ be the Kraus representation of $\mathcal{E}$, where $K_i$'s are Kraus operators. Then the dual map $\mathcal{E}^*$ can be expressed as $\mathcal{E}^*(h)=\sum_i K_i^\dagger hK_i$.} of $\mathcal{E}$ \cite{Zhang2016PRA}; that is, $\mathcal{E}$ and $\mathcal{E}^*$ satisfy the relation $\tr[h_1\mathcal{E}(h_2)]=\tr[\mathcal{E}^*(h_1)h_2]$. It is a general property that $\mathcal{E}^*$ is invertible if and only if $\mathcal{E}$ is invertible. $\mathcal{E} ^{*-1}$ denotes the inverse of $\mathcal{E}^*$. We prove in \cref{appsec_legaldelta} that $\delta$ defined in Eq.~(\ref{eq_construct_delta}) is indeed a legitimate influence operator.

According to Theorem 1 in Ref.~\cite{TAD20}, the QFI $\mathcal{J}[\overline{X}; \mathcal{E}(\rho)]$ can be evaluated using the influence operator:
\begin{eqnarray}\label{QFI-formula}
    1\big/\mathcal{J}[\overline{X}; \mathcal{E}(\rho)]=\min_{h\in\mathcal{T}_{\mathcal{E}(\rho)}^\bot}\norm{\delta-h}_{\mathcal{E}(\rho)}^2,
\end{eqnarray}
which is the convenient formula we seek. We schematically show the concepts introduced above in Fig.~\ref{fig_Hilbert_rho}.

\section{Optimal measurement} \label{sec_optstr}

To find the optimal measurement, we now convert formula~(\ref{QFI-formula}) into an SDP. We do this in the following three steps.

\subsection{Step 1}

We introduce an auxiliary subspace of Hermitian operators
\begin{equation}\label{def-A}
    \mathcal{A} \coloneqq \left\{ h:\mathrm{tr}\left( \rho h \right) =0,~~\mathrm{tr}\left( \frac{\partial \rho}{\partial \theta _i}h \right) =0,~~i=1,...,q \right\} .
\end{equation}
A useful result is that $\mathcal{A}$ can be explicitly characterized as
\begin{equation} \label{eq_cha_mcA}
    \mathcal{A} =\left\{ \mathcal{Q} \left( h \right) :h~~\textrm{is Hermitian} \right\} ,
\end{equation}
with $\mathcal{Q}:=\mathrm{id}-\mathcal{P}$. Here $\mathrm{id}$ denotes the identity map and $\mathcal{P}$ is defined below Eq.~(\ref{Y}). Let us prove the above result. To show that $\mathcal{Q}(h)$ satisfies the two equalities in Eq.~(\ref{def-A}) for any Hermitian operator $h$, we resort to the equality
\begin{eqnarray}
    \mathcal{P}(\rho)=\rho,
\end{eqnarray}
which follows from Eq.~(\ref{symmetric-structure}) and the definition of $\mathcal{P}$. We have
\begin{equation} \label{eq_rhoQh}
    \mathrm{tr}\left[ \rho \mathcal{Q} \left( h \right) \right]= \mathrm{tr}\left( \rho h \right) -\mathrm{tr}\left[ \rho \mathcal{P} \left( h \right) \right] =0,
\end{equation}
where we have used the fact that $\mathrm{tr}\left[ \rho \mathcal{P} \left( h \right) \right]=\mathrm{tr}\left[ \mathcal{P} \left(\rho \right) h \right]$. An immediate consequence of Eq.~(\ref{eq_rhoQh}) is that
\begin{eqnarray}
    \mathrm{tr}\left[\frac{\partial \rho}{\partial\theta_i} \mathcal{Q} \left( h \right) \right]= \frac{\partial}{\partial\theta_i}\mathrm{tr}\left[\rho \mathcal{Q} \left( h \right) \right]=0,
\end{eqnarray}
that is, $\mathcal{Q}(h)$ also satisfies the second equality in Eq.~(\ref{def-A}).
On the other hand, given an operator $h$ satisfying the two equalities in Eq.~(\ref{def-A}), we can decompose it as
\begin{eqnarray}
    h=\mathcal{P} \left( h \right) +\mathcal{Q} \left( h \right).
\end{eqnarray}
Note that $\mathcal{Q}(h)$ satisfies the two equalities in Eq.~(\ref{def-A}), as just proved. We have that $\mathcal{P} \left( h \right) =h-\mathcal{Q} \left( h \right) $, which is a linear combination of $h$ and $\mathcal{Q}(h)$, also satisfies these two equalities. It follows that (see \cref{appsec_proofPhZero} for the proof)
\begin{eqnarray}
    \mathcal{P}(h)=0.
\end{eqnarray}
Therefore,
\begin{eqnarray}
    h=\mathcal{Q}(h),
\end{eqnarray}
implying that the $h$ belongs to the set defined by Eq.~(\ref{eq_cha_mcA}).

\subsection{Step 2}

We show that the space $\mathcal{T} _{\mathcal{E} \left( \rho \right)}^{\bot}$ can be characterized as
\begin{equation} \label{eq_connection_Trhobot}
    \mathcal{T} _{\mathcal{E} \left( \rho \right)}^{\bot}=\{\mathcal{E} ^{*-1}\left( h \right): h\in \mathcal{A}\}.
\end{equation}
Let $h\in \mathcal{A}$, i.e., $h$ satisfies the two equalities in Eq.~(\ref{def-A}).
We have
\begin{equation} \label{appeq_zrhoproof}
    \mathrm{tr}\left[ \mathcal{E} \left( \rho \right) \mathcal{E} ^{*-1}\left( h \right) \right] =\mathrm{tr}\left[ \rho \mathcal{E}^* \left( \mathcal{E} ^{*-1}\left( h \right) \right) \right] =\mathrm{tr}\left( \rho h \right) =0,
\end{equation}
indicating that $\mathcal{E} ^{*-1}\left( h \right)\in\mathcal{Z} _{\mathcal{E} \left( \rho \right)}$. Besides, simple algebra shows
\begin{equation}\label{eq31}
    \langle \mathcal{E} ^{*-1}\left( h \right) ,S_i\rangle _{\mathcal{E} \left( \rho \right)}=\mathrm{tr}\left[  \frac{S_i\mathcal{E} \left( \rho \right) +\mathcal{E} \left( \rho \right) S_i}{2} \mathcal{E} ^{*-1}\left( h \right)\right],
\end{equation}
which, in conjunction with Eq.~(\ref{SLD}),  leads to
\begin{equation}\label{eq32}
    \langle \mathcal{E} ^{*-1}\left( h \right) ,S_i\rangle _{\mathcal{E} \left( \rho \right)}=\mathrm{tr}\left[\frac{\partial}{\partial \theta _i}\mathcal{E} \left( \rho \right) \mathcal{E} ^{*-1}\left( h \right)\right].
\end{equation}
Using $\frac{\partial}{\partial \theta _i}\mathcal{E} \left( \rho \right) = \mathcal{E}(\frac{\partial\rho}{\partial\theta_i})$, we can rewrite the right-hand side of Eq.~(\ref{eq32}) as
\begin{eqnarray}
    \mathrm{tr}\left[\frac{\partial}{\partial \theta _i}\mathcal{E} \left( \rho \right) \mathcal{E} ^{*-1}\left( h \right)\right]
    =
    \mathrm{tr}\left( \frac{\partial\rho}{\partial \theta _i} h \right).
\end{eqnarray}
Further, from the second equality in Eq.~(\ref{def-A}), it follows that
\begin{eqnarray}
    \langle \mathcal{E} ^{*-1}\left( h \right) ,S_i\rangle _{\mathcal{E} \left( \rho \right)}=0.
\end{eqnarray}
So $\mathcal{E} ^{*-1}\left( h \right) $ belongs to $\mathcal{T} _{\mathcal{E} \left( \rho \right)}^{\bot}$ for any $h\in\mathcal{A}$. It remains to show that for any $\tilde{h}\in \mathcal{T} _{\mathcal{E} \left( \rho \right)}^{\bot}$, there exists a $h\in\mathcal{A}$ such that
\begin{eqnarray}\label{eq33}
    \tilde{h}=\mathcal{E}^{*-1}(h).
\end{eqnarray}
Recall that $\tilde{h}$ satisfies two equalities [see Eqs.~(\ref{def-Z}) and (\ref{eq_def_t_Erhobot})]
\begin{eqnarray}\label{eq34}
    \mathrm{tr}\left[ \mathcal{E} \left( \rho \right) \tilde{h} \right] =0,
\end{eqnarray}
and
\begin{eqnarray}\label{eq35}
    \langle \tilde{h}, S_i\rangle_{\mathcal{E}(\rho)}=0.
\end{eqnarray}
Resorting to the same reasoning as in the derivations of Eqs.~(\ref{appeq_zrhoproof}) and (\ref{eq32}), we can respectively rewrite Eqs.~(\ref{eq34}) and (\ref{eq35}) as
\begin{eqnarray}\label{eq36}
    \tr\left[\rho\mathcal{E}^*(\tilde{h})\right]=0,
\end{eqnarray}
and
\begin{equation} \label{eq37}
    \mathrm{tr}\left[ \frac{\partial \rho}{\partial \theta _i}\mathcal{E} ^*\left( \tilde{h} \right) \right] =0.
\end{equation}
Comparing Eqs.~(\ref{eq36}) and (\ref{eq37}) with the two equalities in Eq.~(\ref{def-A}), we see that $\mathcal{E} ^*\left( \tilde{h} \right)$ belongs to $\mathcal{A}$. That is, $\mathcal{E} ^*\left( \tilde{h} \right)=h$ for some $h\in\mathcal{A}$, which is equivalent to Eq.~(\ref{eq33}).

\subsection{Step 3}

We specify the SDP for finding the optimal measurement. Inserting \cref{eq_cha_mcA,eq_connection_Trhobot} into Eq.~(\ref{QFI-formula}) gives
\begin{align}
    1/\mathcal{J} \left[ \overline{X};\mathcal{E} \left( \rho \right) \right] =\min_{h} \left\| \delta -\mathcal{E} ^{*-1}\left[ \mathcal{Q} \left( h \right) \right] \right\| _{\mathcal E(\rho)}^{2}. \label{eq_reform2}
\end{align}
We introduce the observable
\begin{equation} \label{eq_defYh}
    Y_h\coloneqq \mathcal{E} ^{*-1}\left[ X-\mathcal{Q} \left( h \right) \right],
\end{equation}
which satisfies
\begin{equation}\label{exp-Y-X}
    \mathrm{tr}\left[ \mathcal{E} \left( \rho \right) Y_h \right] =\mathrm{tr}\left[ \rho \mathcal{E} ^*\left( Y_h \right) \right] =\mathrm{tr}\left( \rho X \right) -\mathrm{tr}\left[ \rho \mathcal{Q} \left( h \right) \right] =\overline{X}.
\end{equation}
That is, the expectation value of $Y_h$ in the corrupted state $\mathcal{E}(\rho)$ is equal to the expectation value of $X$ in $\rho$. Noting that $\delta =\mathcal{E} ^{*-1}\left( X \right) -\overline{X} \mathbb{1}$ [see Eq.~(\ref{eq_construct_delta})], we can rewrite
\cref{eq_reform2} as
\begin{equation} \label{eq_var}
    1/\mathcal{J} \left[ \overline{X};\mathcal{E} \left( \rho \right) \right]=\underset{h}{\min}\,\,\mathrm{tr}\left[ \mathcal{E} \left( \rho \right) \left( Y_h-\overline{X}\mathbb{1} \right) ^2 \right].
\end{equation}
It is important to notice that the term $\mathrm{tr}\left[ \mathcal{E} \left( \rho \right) \left( Y_h-\overline{X}\mathbb{1} \right) ^2 \right]$ appearing in Eq.~(\ref{eq_var}) is simply the quantum uncertainty of $Y_h$ in the state $\mathcal{E}(\rho)$; that is,
\begin{eqnarray}
    1/\mathcal{J} \left[ \overline{X};\mathcal{E} \left( \rho \right) \right]=\min_{h}\left(\Delta Y_h\right)_{\mathcal{E}(\rho)}^2,
\end{eqnarray}
where we use $\left(\Delta Y_h\right)_{\mathcal{E}(\rho)}^2$ to denote the quantum uncertainty of $Y_h$ in $\mathcal{E}(\rho)$.
Lastly, we reformulate Eq.~(\ref{eq_var}) as the SDP:
\begin{subequations} \label{eq_sdpform}
    \begin{align}
        1/\mathcal{J} \left[ \overline{X};\mathcal{E} \left( \rho \right) \right]=\underset{\Lambda,\,\,h}{\min}\quad & \mathrm{tr}\left[ \mathcal{E} \left( \rho \right) \Lambda \right]
        \\
        \mathrm{s}.\mathrm{t}.\quad                                                                                   & \left[ \begin{matrix}
                                                                                                                                       \Lambda                     & Y_h - \overline{X}\mathbb 1 \\
                                                                                                                                       Y_h - \overline{X}\mathbb 1 & \mathbb{1}                  \\
                                                                                                                                   \end{matrix} \right] \ge 0.
    \end{align}
\end{subequations}
The correctness of this reformulation can be verified by noting that the constraint in \cref{eq_sdpform} can equivalently be expressed as $\Lambda \ge \left( Y_h-\overline{X}\mathbb{1} \right) ^2$ according to the Schur complement condition for positive semidefiniteness \cite{HJ12}. We cast Eq.~(\ref{eq_sdpform}) in the canonical form of an SDP in \cref{appsec_proof_eq_sdpform_SDP}.

We now summarize the results obtained so far as a theorem.

\begin{theorem} \label{theorem1}
    Let the symmetric structures of $\rho$ be described by a finite or compact Lie group $G$. The optimal measurement to retrieve the information about $\overline{X}=\tr(\rho X)$ from the corrupted state $\mathcal{E}(\rho)$ is the projective measurement of the observable $Y_{h_0}$ defined in \cref{eq_defYh}, where $h_0$ minimizes the SDP in \cref{eq_sdpform}. The expectation value of $Y_{h_0}$ in $\mathcal{E}(\rho)$ equals to the expectation value of $X$ in $\rho$. Moreover, the inverse of the QFI $\mathcal{J} \left[ \overline{X};\mathcal{E} \left( \rho \right) \right]$ equals to the quantum uncertainty of $Y_{h_0}$ in $\mathcal{E}(\rho)$.
\end{theorem}

We clarify that the optimal solution $h_0$ may depend on the state $\rho$ in general. Physically, this means that a refined knowledge of $\rho$ may be required in order to implement the projective measurement of $Y_{h_0}$ in practice. It should be mentioned that such an unpleasant dependence is a common feature of quantum metrological protocols \cite{PAR09} rather than being exclusive to our study. Below, we eliminate the dependence of $h_0$ on $\rho$ by focusing on the noise models that are covariant under $G$.

\section{Covariant noise models} \label{sec_covchannel}
Let us now consider the setting that $\mathcal{E}$ is covariant under the group $G$, which is of relevance in a plethora of physical contexts as mentioned in the introduction \cite{BRS07,WWW52,BW03,MS14,PCB16,ZYHT17,GS08,CG19}.
Formally, $\mathcal{E}$ is said to be covariant with respect to $G$ if
\begin{equation} \label{eq_covariant_def}
    \mathcal{E} \left( U_ghU_{g}^{\dagger} \right) =U_g\mathcal{E} \left( h \right) U_{g}^{\dagger},
\end{equation}
for all $g\in G$ and Hermitian operators $h$ \cite{MS14,PCB16,ZYHT17}. In what follows, we show how to explicitly determine the optimal measurement of $Y_{h_0}$ for these covariant noise models.

We first show that $\mathcal{P}$ and $\mathcal{E}^{*-1}$ are exchangeable, that is,
\begin{eqnarray}\label{exc-P-E}
    \mathcal{E}^{*-1}(\mathcal{P}(h))=\mathcal{P}(\mathcal{E}^{*-1}(h)),
\end{eqnarray}
for any $h$. To see why Eq.~(\ref{exc-P-E}) holds, we can sum both sides of \cref{eq_covariant_def} over the group elements $g$ in $G$, and obtain
\begin{equation} \label{eq_PE_commute}
    \mathcal{E} \left[ \mathcal{P} \left( h \right) \right] =\mathcal{P} \left[ \mathcal{E} \left( h \right) \right],
\end{equation}
for all $h$. That is, the two maps $\mathcal{P}$ and $\mathcal{E}$ are exchangeable. This implies that $\mathcal{P}$ and $\mathcal{E}^{*-1}$ are exchangeable, which can be verified by resorting to the matrix representation of $\mathcal{E}$, $\mathcal{P}$, and $\mathcal{E}^{*-1}$. Indeed,
let $\left\{ H _i \right\} $ denote a Hermitian basis. In this basis, the linear map $\mathcal{E}$ can be represented by the matrix $A_\mathcal{E}$ whose $ij$th element is given by
\begin{eqnarray}\label{matrix-A}
    A_{\mathcal{E} ,ij}=\mathrm{tr}\left[ H _i\mathcal{E} \left( H _j \right) \right].
\end{eqnarray}
Then, by the cyclic property of the trace, we have that
\begin{eqnarray}
    A_{\mathcal{E},ij} = \mathrm{tr}\left[ H _i\mathcal{E} \left( H _j \right) \right] =\mathrm{tr}\left[ \mathcal{E} ^*\left( H_i \right) H_j \right] =A_{\mathcal{E} ^*,ji},
\end{eqnarray}
which implies that $A_{\mathcal{E}}^{T}=A_{\mathcal{E} ^*}$.
Additionally, $A_{\mathcal{P}}$ is symmetric since
\begin{eqnarray}
    A_{\mathcal{P} ,ij}=\mathrm{tr}\left[ H _i\mathcal{P} \left( H _j \right) \right]  =\mathrm{tr}\left[ \mathcal{P} \left( H _i \right) H _j \right]=A_{\mathcal{P} ,ji}.
\end{eqnarray}
Combining these facts, we have that the commutativity $\left[ A_{\mathcal{P}},A_{\mathcal{E}} \right] =0$, which follows from Eq.~(\ref{eq_PE_commute}), implies that $\left[ A_{\mathcal{P}}^{T},A_{\mathcal{E}}^{T} \right] =\left[ A_{\mathcal{P}},A_{\mathcal{E} ^*} \right] =0$. Furthermore, using the identity $A_{\mathcal{E} ^{*-1}}=A_{\mathcal{E} ^*}^{-1}$, we conclude that $\left[ A_{\mathcal{P}},A_{\mathcal{E} ^{*-1}} \right] =0$ as well, that is, $\mathcal{P}$ and $\mathcal{E}^{*-1}$ are exchangeable. Noting that
\begin{eqnarray}
    \mathcal{Q}=\textrm{id}-\mathcal{P},
\end{eqnarray}
we deduce from Eq.~(\ref{exc-P-E}) that $\mathcal{Q}$ and $\mathcal{E}^{*-1}$ are exchangeable, too.

We then show that $\mathcal{P}$ is idempotent and Hermitian with respect to the inner product in Eq.~(\ref{inner-product}), that is, $\mathcal{P}$ satisfies
\begin{equation}\label{fact-iii-1}
    \mathcal{P}\left(\mathcal{P}(h_1)\right)=\mathcal{P}(h_1),
\end{equation}
and
\begin{eqnarray}\label{fact-ii}
    \langle \mathcal{P}(h_1), h_2\rangle _{\mathcal{E}(\rho)}=\langle h_1, \mathcal{P}(h_2)\rangle _{\mathcal{E}(\rho)},
\end{eqnarray}
for any operators $h_1$ and $h_2$. Equation (\ref{fact-iii-1}) follows directly from the fact that $\mathcal{P}(h_1)$ is invariant under the action of $U_g$, i.e., $U_g\mathcal{P}(h_1)U_g^\dagger=\mathcal{P}(h_1)$.
To verify Eq.~(\ref{fact-ii}), we note that
\begin{equation}
    \mathrm{tr}\left[ \mathcal{E} \left( \rho \right) \mathcal{P} \left( h_1 \right) h_2 \right] =\frac{1}{\abs{G}}\sum_{g\in G} \mathrm{tr}\left[ U_{g}^{\dagger}\mathcal{E} \left( \rho \right) U_gh_1U_{g}^{\dagger}h_2U_g \right],
\end{equation}
which leads to
\begin{align}\label{eq_Ph2h1_h2Ph1}
    \mathrm{tr}\left[ \mathcal{E} \left( \rho \right) \mathcal{P} \left( h_1 \right) h_2 \right] =\mathrm{tr}\left[ \mathcal{E} \left( \rho \right) h_1\mathcal{P} \left( h_2\right) \right],
\end{align}
as $\mathcal{E}$ is covariant under $G$ and $\rho$ is symmetric.
Analogously, we have
\begin{equation} \label{eq_h1Ph2_Ph1h2}
    \mathrm{tr}\left[ \mathcal{E} \left( \rho \right) h_2\mathcal{P} \left( h_1 \right) \right] =\mathrm{tr}\left[ \mathcal{E} \left( \rho \right) \mathcal{P} \left( h_2 \right) h_1 \right] .
\end{equation}
Then Eq.~(\ref{fact-ii}) follows from summing the two sides of \cref{eq_Ph2h1_h2Ph1,eq_h1Ph2_Ph1h2}. A direct consequence of Eqs.~(\ref{fact-iii-1}) and (\ref{fact-ii}) is
\begin{eqnarray}\label{fact-iii}
    \langle \mathcal{P}(h_1), \mathcal{Q}(h_2) \rangle_{\mathcal{E}(\rho)}=0,
\end{eqnarray}
which can be verified by noting that $\mathcal{P}(\mathcal{Q}(h))=\mathcal{P}(h)-\mathcal{P}(\mathcal{P}(h))=0$.

Let us now determine $Y_{h_0}$. To do this, we use the equality
\begin{eqnarray}
    \mathrm{tr}\left[ \mathcal{E} \left( \rho \right) \left( Y_h-\overline{X}\mathbb{1} \right) ^2 \right]=\tr\left[\mathcal{E}(\rho)Y_h^2\right]-\overline{X}^2
\end{eqnarray}
to
rewrite Eq.~(\ref{eq_var}) as
\begin{equation} \label{eq_var-re}
    1/\mathcal{J} \left[ \overline{X};\mathcal{E} \left( \rho \right) \right]=\min_{h}\tr\left[\mathcal{E}(\rho)Y_h^2\right]-\overline{X}^2.
\end{equation}
Note that $Y_h$ can be expressed as
\begin{eqnarray}\label{sec-model-1}
    Y_h=\mathcal{E}^{*-1}(\mathcal{P}(X))+\mathcal{E}^{*-1}(\mathcal{Q}(X-h)).
\end{eqnarray}
Exchanging $\mathcal{E}^{*-1}$ with $\mathcal{P}$ and $\mathcal{Q}$ in Eq.~(\ref{sec-model-1}), we have that
\begin{eqnarray}\label{sec-model-2}
    Y_h=\mathcal{P}(\mathcal{E}^{*-1}(X))+\mathcal{Q}(\mathcal{E}^{*-1}(X-h)).
\end{eqnarray}
Inserting Eq.~(\ref{sec-model-2}) into Eq.~(\ref{eq_var-re}) and using Eq.~(\ref{fact-iii}), we obtain
\begin{eqnarray}\label{eq_threeterms_middle_min}
    1/\mathcal{J} \left[ \overline{X};\mathcal{E} \left( \rho \right) \right]&=&\underset{h}{\min}\left\| \mathcal{Q} \left( \mathcal{E} ^{*-1}\left( X-h \right) \right) \right\| _{\mathcal{E} \left( \rho \right)}^{2}\nonumber\\
    &+&\left\| \mathcal{P} \left( \mathcal{E} ^{*-1}\left( X \right) \right) \right\| _{\mathcal{E} \left( \rho \right)}^{2}-\overline{X}^2.
\end{eqnarray}
Apparently, $h=X$ attains the minimum and $Y_{h_0}=\mathcal{P}(\mathcal{E}^{*-1}(X))=\mathcal{E}^{*-1}(\mathcal{P}(X))$. We therefore arrive at the following theorem:

\begin{theorem} \label{th_covariant}
    The observable $Y_{h_0}$ is $\mathcal{E} ^{*-1}\left( \mathcal{P} \left( X \right) \right)$ or equivalently $\mathcal{P} \left( \mathcal{E} ^{*-1}\left( X \right) \right) $ when the noise model $\mathcal{E}$ is covariant under $G$.
\end{theorem}

We see that the observable $\mathcal{E} ^{*-1}\left( \mathcal{P} \left( X \right) \right)$ is independent of $\rho$, which means that the aforementioned dependence issue is eliminated in Theorem \ref{th_covariant}. We clarify that Theorem \ref{th_covariant} holds for any covariant quantum operation $\mathcal{E}$ and any choice of $X$. It is worth noting that a special covariant quantum operation is $\mathcal{E}=\textrm{id}$, which corresponds to the noiseless situation considered in Ref.~\cite{ZT24}. We easily deduce from Theorem \ref{th_covariant} that $Y_{h_0}=\mathcal{P}(X)$ in this situation, which is one of the key findings of Ref.~\cite{ZT24}. Besides, to better digest the result in Theorem \ref{th_covariant}, we provide an intuitive understanding of this result in \cref{appsec_intuitive}.

\section{Illustrative application} \label{sec_application}
To demonstrate the usefulness of our results, we now consider the setting that $\rho$ is an $n$-qubit state whose symmetric structures are described by $G=\{P_{\pi}, \pi\in S_n\}$. Here $\pi$ labels the permutations in the symmetric group $S_n$ and $P_{\pi}$ is the unitary representation of $\pi$, defined by
\begin{equation}
    P_{\pi}|\psi _1\rangle \otimes \cdots \otimes |\psi _n\rangle =|\psi _{\pi ^{-1}\left( 1 \right)}\rangle \otimes \cdots \otimes |\psi _{\pi ^{-1}\left( n \right)}\rangle,
\end{equation}
with $|\psi _i\rangle \in \mathbb C^2$.
Such states naturally arise in multipartite experiments \cite{KST07,WSK08,WKK09,PCT09,KWO10,EMKZ18,LZM21}. A well-known example is the Dicke states
\begin{equation} \label{eq_dicke}
    |D_{n}^{\left( l \right)}\rangle =\left( \begin{array}{c}
        n \\
        l \\
    \end{array} \right) ^{-1/2}\sum_{x\in\{0, 1\}^n,~\textrm{wt}(x)=l}\ket{x},
\end{equation}
where $\textrm{wt}(x)$ is the number of ones in $x$, e.g., $\textrm{wt}(x)=1$ when $x=010$.

Motivated by the fact that dephasing is one of the dominant types of noise encountered in experiments \cite{PRG19,ZBH19,ZZW22}, we set $\mathcal{E}$ to be the dephasing noise acting independently on the $n$ qubits. Specifically, the dephasing noise acting on a single qubit is described by the channel
\begin{equation}
    \mathcal{E} _p\left( \rho \right) =\left( 1-\frac{p}{2} \right) \rho +\frac{p}{2}\sigma _z\rho \sigma _z,
\end{equation}
where $p$ quantifies the noise strength. Here we assume that $0\le p <1$, since the dephasing channel is not invertible at $p=1$. The noise model $\mathcal{E}$ can be described as
\begin{equation}
    \mathcal{E} =\mathcal{E} _{p}^{\otimes n}.
\end{equation}
A direct calculation shows that the map $\mathcal{E}^{*-1}$ reads
\begin{equation}
    \mathcal{E} ^{*-1}=\mathcal{E} _{p/\left( p-1 \right)}^{\otimes n}.
\end{equation}
Notably, $\mathcal{E}$ is covariant under $G$, which implies that \cref{th_covariant} can be applied in the setting under consideration.

\begin{figure}
    \centering
    \includegraphics[width=\linewidth]{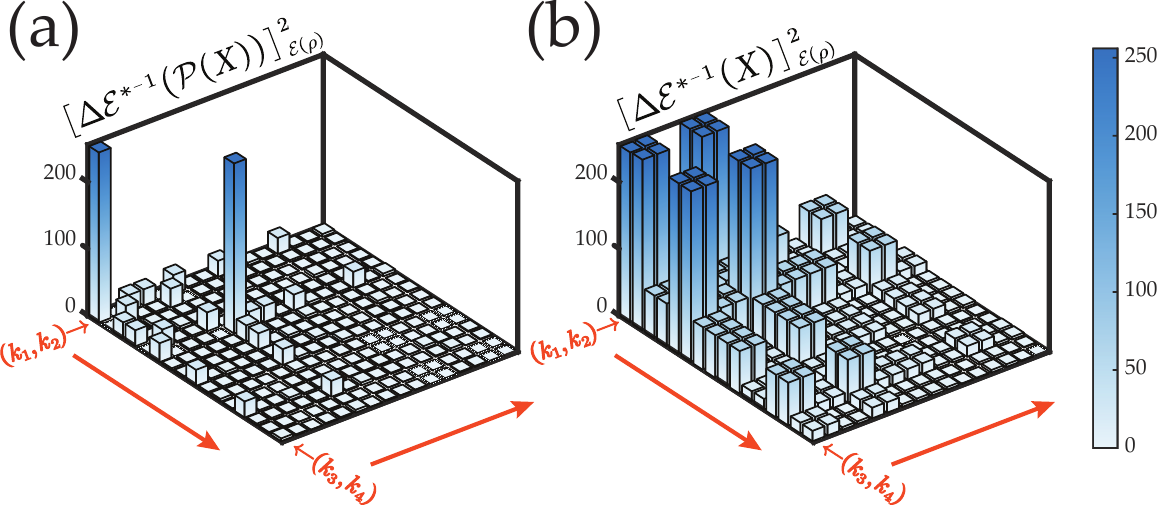}
    \caption{Illustration of the quantum uncertainties associated with (a) the optimal measurement identified in the present work and (b) the customary measurement studied in Ref.~\cite{WSU10}.
    Here, $\rho$ is set to be the four-qubit Dicke state $\ket{D_4^{(2)}}$.
    The observables under consideration are the Pauli observables $X_{ k_1,k_2,k_3,k_4}$ in Eq.~(\ref{eq_pauli_obs}). The height of the bins represents the quantum uncertainties of (a) $\mathcal{E} ^{*-1}\left( \mathcal{P} \left( X \right) \right) $ and (b) $\mathcal{E} ^{*-1}\left( X \right)$ in $\mathcal{E}(\rho)$.}
    \label{fig_app_pauli}
\end{figure}

To demonstrate the superiority of the projective measurement of the observable $\mathcal{E} ^{*-1}\left( \mathcal{P} \left( X \right) \right)$, which is the optimal measurement according to \cref{th_covariant},
we would like to compare it with the projective measurement of the observable $\mathcal{E}^{*-1}(X)$. The latter measurement, referred to as the customary measurement hereafter, has been studied in Ref.~\cite{WSU10}. It is interesting to note that
\begin{eqnarray}
    \tr[\mathcal{E}(\rho)\mathcal{E}^{*-1}(X)]=\tr(\rho X),
\end{eqnarray}
that is, the expectation value of $\mathcal{E}^{*-1}(X)$ in $\mathcal{E}(\rho)$ is equal to $\overline{X}$. Hence, both the optimal measurement identified here and the customary measurement can be employed to measure $\overline{X}$. The interesting difference between them is that
\begin{eqnarray}
    [\Delta\mathcal{E}^{*-1}(\mathcal{P}(X))]_{\mathcal{E}(\rho)}^2\leq [\Delta\mathcal{E}^{*-1}(X)]_{\mathcal{E}(\rho)}^2,
\end{eqnarray}
that is,
the quantum uncertainty of $\mathcal{E} ^{*-1}\left( \mathcal{P} \left( X \right) \right)$ in $\mathcal{E}(\rho)$ is generally smaller than the quantum uncertainty of $\mathcal{E}^{*-1}(X)$.

\begin{figure}
    \centering
    \includegraphics[width=\linewidth]{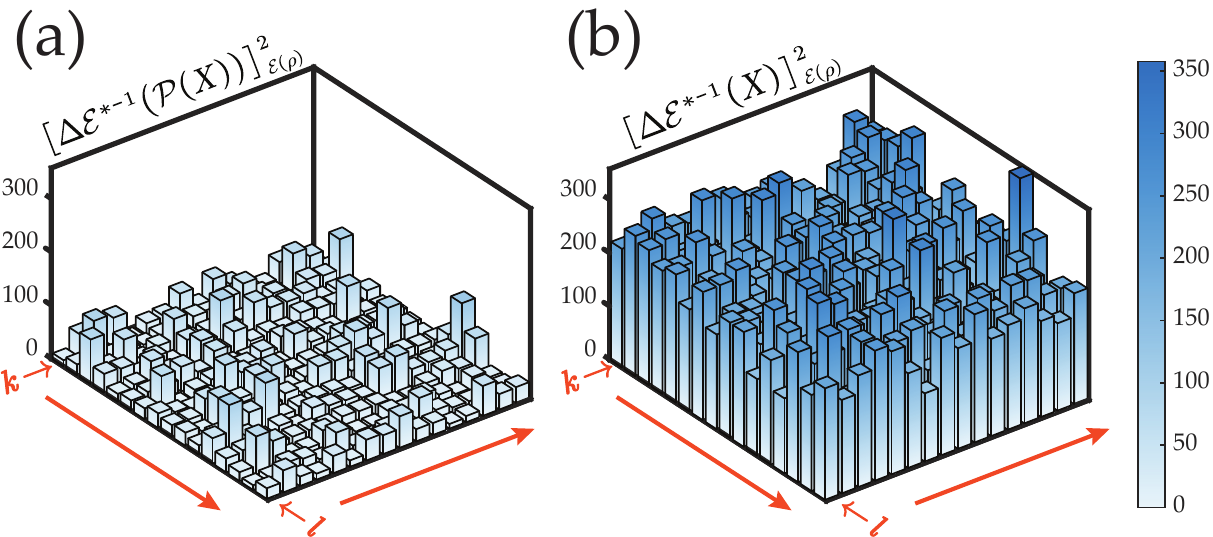}
    \caption{Illustration of the two quantum uncertainties for $256$ randomly generated observables. Here, $\rho$ is set to be the Dicke state $\ket{D_4^{(2)}}$, too. The observables are indexed as $X_{ k,l }$ with $k$ and $l$ ranging from $1$ to $16$. The height of the bins represents the quantum uncertainty of (a) $\mathcal{E} ^{*-1}\left( \mathcal{P} \left( X \right) \right) $ and (b) $\mathcal{E} ^{*-1}\left( X \right)$ in $\mathcal{E}(\rho)$.}
    \label{fig_app_random}
\end{figure}

To explicitly see the difference, we consider the task of measuring the expectation values of Pauli observables
\begin{equation} \label{eq_pauli_obs}
    X_{k_1,k_2,\cdots,k_n}=\sigma _{k_1}\otimes \sigma _{k_2}\otimes \cdots \otimes \sigma _{k_n},
\end{equation}
where $\sigma_{k_i}\in\left\{\sigma_x,\sigma_y,\sigma_z,\sigma_I\right\}$ with $\sigma_I=\mathbb 1$. Figure \ref{fig_app_pauli} shows the values of the two quantum uncertainties when $\rho$ is the four-qubit Dicke state $\ket{D_4^{(2)}}$. The labels $(k_1, k_2)$ and $(k_3, k_4)$, which appear aside the two horizontal axes in Fig.~\ref{fig_app_pauli}, correspond to the observable $X_{k1, k2, k3, k4}$ in Eq.~(\ref{eq_pauli_obs}).
As can be seen from \cref{fig_app_pauli}, $[\Delta\mathcal{E}^{*-1}(\mathcal{P}(X))]_{\mathcal{E}(\rho)}^2$ is significantly smaller than $[\Delta\mathcal{E}^{*-1}(X)]_{\mathcal{E}(\rho)}^2$ for most of the $256$ Pauli observables. Specifically,
we find that the ratio
\begin{eqnarray}\label{ratio-Pauli}
    [\Delta\mathcal{E}^{*-1}(\mathcal{P}(X))]_{\mathcal{E}(\rho)}^2\big/[\Delta\mathcal{E}^{*-1}(X)]_{\mathcal{E}(\rho)}^2\leq 0.37,
\end{eqnarray}
for all the Pauli observables except for the trivial ones, $\sigma _{\alpha}^{\otimes 4}$, $\alpha=I, x, y, z$, which are invariant under $\mathcal{P}$, i.e., $\mathcal{P}(\sigma _{\alpha}^{\otimes 4})=\sigma _{\alpha}^{\otimes 4}$. We see from Eq.~(\ref{ratio-Pauli}) that the optimal measurement, when used to measure $\overline{X}$ for these observables, requires only $37\%$ of the copies of $\mathcal{E}(\rho)$ compared with the customary measurement to achieve the same precision.


To demonstrate that the superiority of our measurement is not exclusive to the 256 Pauli observables considered above, we further randomly generate 256 new observables, $X_{k,l}$, with $k,l=1,2,\cdots, 16$. Here, each observable is produced by randomly generating a complex matrix $C$ and then setting $X$ to be $\left( C+C^{\dagger} \right) /2$. Figure \ref{fig_app_random} shows the two quantum uncertainties for the randomly generated observables, where $\rho$ is set to be the Dicke state $\ket{D_4^{(2)}}$, too.
As can be seen from this figure, the quantum uncertainty associated with the optimal measurement is significantly smaller than the customary one for all the observables.   Specifically, the ratio reads
\begin{equation}\label{ratio-random}
    0.01 \leq [\Delta\mathcal{E}^{*-1}(\mathcal{P}(X))]_{\mathcal{E}(\rho)}^2\big/[\Delta\mathcal{E}^{*-1}(X)]_{\mathcal{E}(\rho)}^2\leq 0.42,
\end{equation}
with an average value of $0.14$. We have randomly generated multiple sets of $256$ observables and obtained similar results, although the specific numbers appearing in Eq.~(\ref{ratio-random}) may be different when different sets are in question.

\begin{figure}
    \centering
    \includegraphics[width=\linewidth]{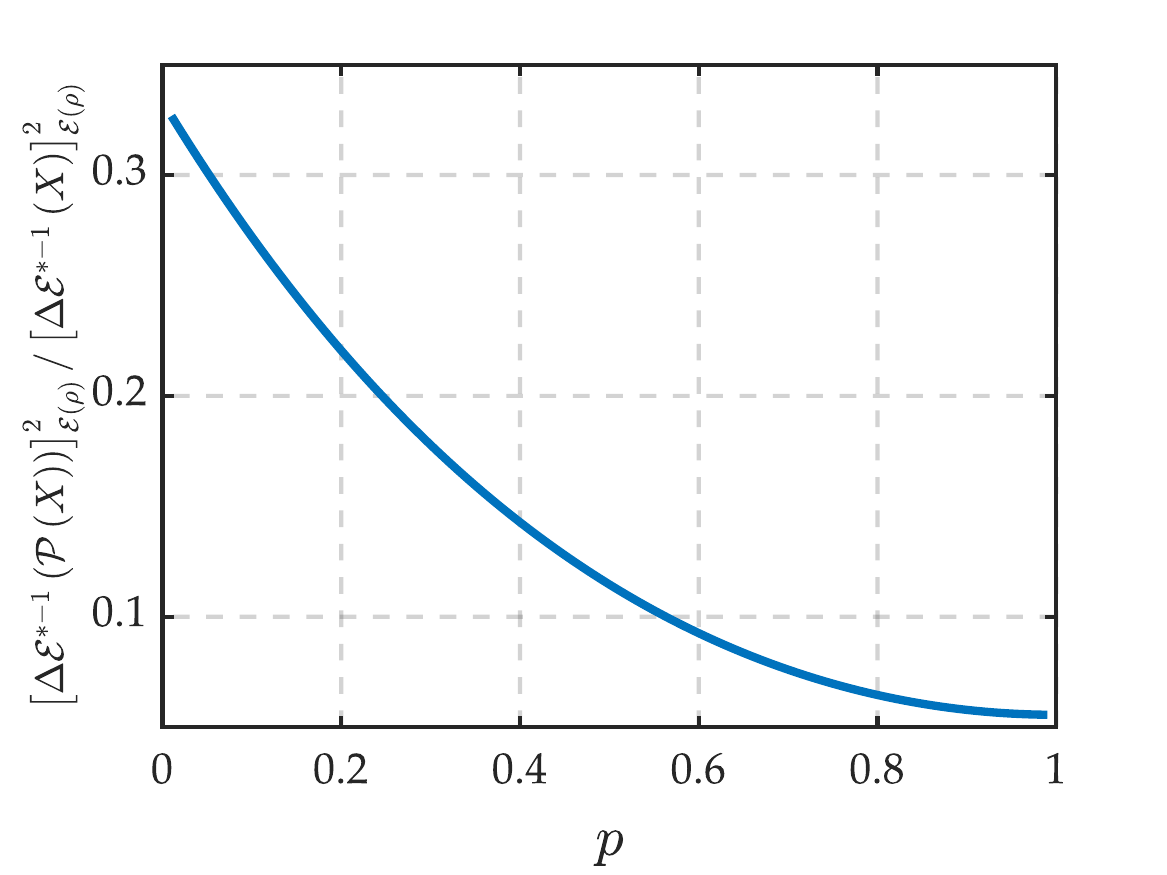}
    \caption{The ratio between the quantum uncertainties of $\mathcal{E} ^{*-1}\left( \mathcal{P} \left( X \right) \right) $ and $\mathcal{E} ^{*-1}\left( X \right) $ as a function of the noise strength $p$.}
    \label{fig_app_diffp}
\end{figure}

\begin{figure}
    \centering
    \includegraphics[width=\linewidth]{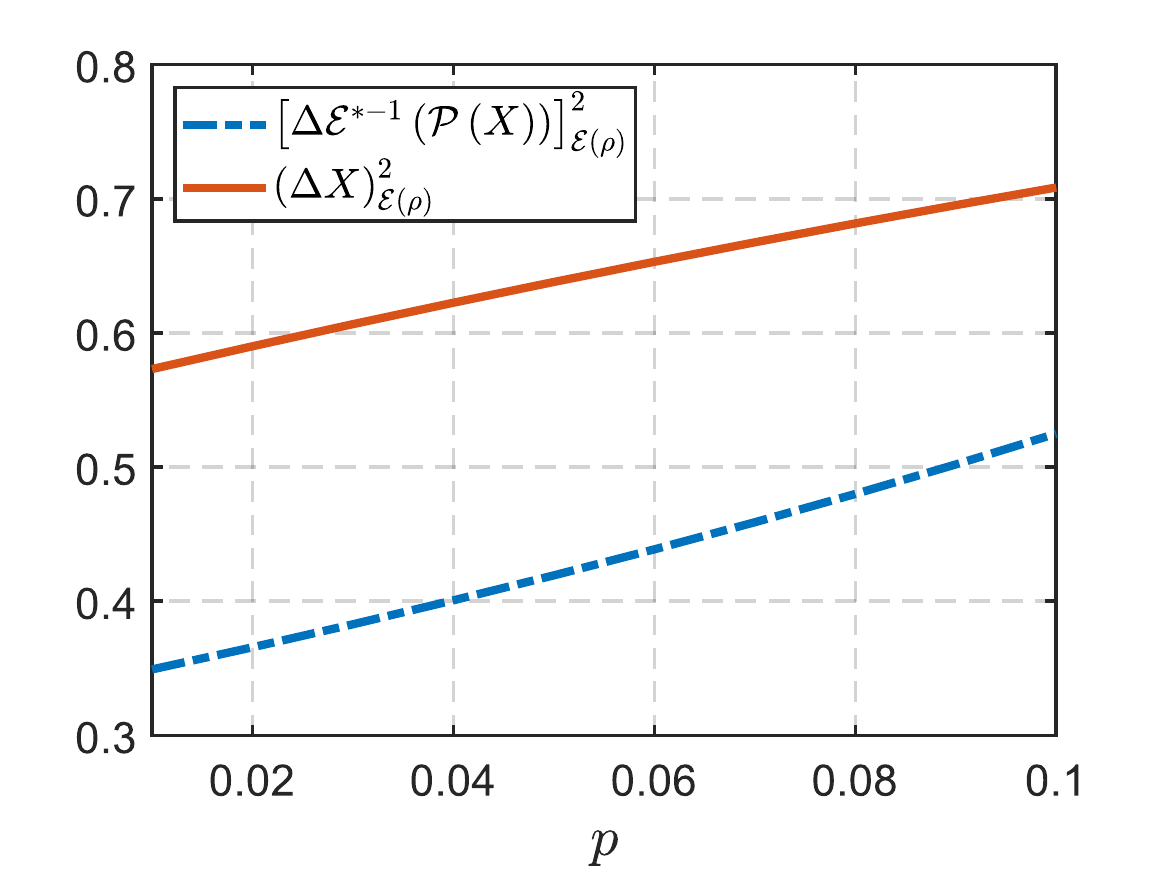}
    \caption{Illustration of the quantum uncertainties of $\mathcal{E} ^{*-1}\left( \mathcal{P} \left( X \right) \right)$ (blue dashed line) and $X$ (red solid line) as functions of the noise strength $p$. Here, $X$ is set to be $\sigma _y\otimes \sigma _y\otimes \sigma _z\otimes \sigma _z$.}
    \label{fig_direct_X}
\end{figure}

Besides, we examine the ratio between the two quantum uncertainties as a function of the noise strength $p$. As can be seen from \cref{fig_app_diffp}, the ratio becomes increasingly small in the course of varying $p$ from $0.01$ to $0.99$.
This means that the superiority of our measurement becomes increasingly significant as the noise strength increases. Moreover, we also numerically examine the performance of the projective measurement of $X$ in the low-noise regime, which may provide an approximately unbiased estimate of $\overline{X}$. The numerical result is presented in \cref{fig_direct_X}, showing that $\left[ \Delta \mathcal{E} ^{*-1}\left( \mathcal{P} \left( X \right) \right) \right] _{\mathcal{E} \left( \rho \right)}^{2}$ is also smaller than $\left( \Delta X \right) _{\mathcal{E} \left( \rho \right)}^{2}$ when $p$ varies from $0.01$ to $0.1$.

\section{Concluding Remarks} \label{sec_conclude_discuss}
We have identified the measurement capable of retrieving the maximum information about the expectation value $\overline{X}$ of an observable $X$ in $\rho$ from the corrupted state $\mathcal{E}(\rho)$. Our general result, presented in Theorem \ref{theorem1}, shows that this measurement is the projective measurement of $Y_{h_0}$, which can be found by solving the SDP in Eq.~(\ref{eq_sdpform}). As the QFI is employed to quantify the information content, Theorem \ref{theorem1} is of fundamental importance and characterizes the optimal sample complexity in estimating $\overline{X}$ according to the quantum Cram\'{e}r-Rao bound \cite{ZT24}.

To eliminate the dependence of $Y_{h_0}$ on $\rho$, we have shown that $Y_{h_0}$ can be explicitly determined for covariant noise models, as stated in Theorem \ref{th_covariant}. This result reduces to one of the key findings in Ref.~\cite{ZT24}, which can find immediate applications due to the relevance of covariant noise models in various contexts. We have demonstrated the usefulness of our result by applying Theorem \ref{th_covariant} to the scenario that involves permutation symmetries.

We clarify that, while the local parameter estimation theory provides powerful tools like the QFI, it also produces the inherent drawback that $Y_{h_0}$ may depend on $\rho$ \cite{PAR09}.
To overcome this drawback, one may resort to the global parameter estimation theory, such as the Bayesian and minimax approaches \cite{Per71,GL95,Hay11,MFD14,RD19,DGG20,GDWB20,RD20,SK20,GD22,BLSM24,Rub24}, which can single out a globally optimal measurement that is independent of $\rho$. It is worth noting that a recent study \cite{ZQZ24} has established a link between the local and global parameter estimation theories, which could be a valuable reference for future efforts to fully resolve the dependence issue.

\begin{acknowledgments}
    We thank the anonymous referee for sharing with us an intuitive understanding of \cref{th_covariant}.
    This work was supported by the National Natural Science Foundation of
    China through Grant Nos.~12275155 and 12174224.
\end{acknowledgments}

\appendix

\section{Parametrization of $\rho$} \label{appsec_rho}

According to the representation theory of groups \cite{Sim97}, the unitary representation $U_g$ can be written as a direct sum of irreducible unitary representations in a certain basis as
\begin{equation}
    U_g=\bigoplus_{\alpha =1}^s{\mathbb{1} _{n_{\alpha}}\otimes U_{\alpha}\left( g \right)},
\end{equation}
where $\mathbb 1_k$ denotes the $k$ by $k$ identity matrix, and $\alpha$ labels the $\alpha$th irreducible representation of $G$ with dimension $d_\alpha$ and multiplicity $n_\alpha$. Consequently, any Hermitian $h$ commutes with all the $U_g$ can be expressed in the same basis as
\begin{equation} \label{appeq_symmetry_form_h}
    h=\bigoplus_{\alpha =1}^s{A_{\alpha}\otimes \mathbb{1} _{d_{\alpha}}},
\end{equation}
where $A_{\alpha}$ is an $n_\alpha\times n_\alpha$ Hermitian matrix. Accordingly, since $[\rho, U_g]=0$ for all $g\in G$, we can explicitly express $\rho$ as
\begin{equation} \label{appeq_parameterization_rho}
    \rho =\bigoplus_{\alpha =1}^s{\left( p_{\alpha}\frac{\mathbb{1} _{n_{\alpha}}}{n_{\alpha}}+\frac{1}{2}\boldsymbol{r}_{\alpha}\cdot \boldsymbol{\lambda }_{\alpha} \right)}\otimes \frac{\mathbb{1} _{d_{\alpha}}}{d_{\alpha}},
\end{equation}
where $\boldsymbol{\lambda }_{\alpha}\coloneqq \left( \lambda _{\alpha ,1},\lambda _{\alpha ,2},\cdots ,\lambda _{\alpha ,n_{\alpha}^{2}-1} \right) $ collectively denotes the generators of the Lie algebra $\mathfrak{s} \mathfrak{u} \left( n_{\alpha} \right)$ that satisfy \cite{Kim03}
\begin{equation} \label{appeq_generator_relation}
    \lambda _{\alpha ,i}^{\dagger}=\lambda _{\alpha ,i},~~\mathrm{tr}\left( \lambda _{\alpha ,i} \right) =0,~~\mathrm{tr}\left( \lambda _{\alpha ,i}\lambda _{\alpha ,j} \right) =2\delta _{ij}.
\end{equation}
Therefore, $\rho$ can be parameterized by the following real parameters
\begin{equation} \label{appeq_theta_parameterize}
    \boldsymbol{\theta }=\left( p_1,p_2,\cdots ,p_{s-1},\boldsymbol{r}_1,\cdots ,\boldsymbol{r}_s \right)
\end{equation}
with $\boldsymbol{r}_{\alpha}=\left( r_{\alpha ,1},\cdots ,r_{\alpha ,n_{\alpha}^{2}-1} \right) $, and we have set the variable $p_s$ to be $p_s=1-\sum_{i=1}^{s-1}p_i$ due to the constraint $\sum_{i=1}^sp_i=1$.

\section{The legitimacy of $\delta$ in Eq.~(\ref{eq_construct_delta}) as an influence operator} \label{appsec_legaldelta}

To verify that the $\delta$ in Eq.~(\ref{eq_construct_delta}) belongs to $\mathcal{Z}_{\mathcal{E}(\rho)}$, we notice that
\begin{eqnarray}
    \tr[\mathcal{E}(\rho)\mathcal{E} ^{*-1}\left( X \right)]=\tr[\mathcal{E}^{-1}(\mathcal{E}(\rho)) X]=\tr(\rho X),
\end{eqnarray}
leading to
\begin{eqnarray}
    \tr[\mathcal{E}(\rho)\delta]=\tr[\mathcal{E}(\rho)\mathcal{E} ^{*-1}\left( X \right)]-\overline{X}=0.
\end{eqnarray}
Further, using the cyclic property of the trace, we have
\begin{eqnarray}\label[step]{exchange-order}
    \tr[\mathcal{E}(\rho)\frac{S_i\delta+\delta S_i}{2}]=\tr[\frac{\mathcal{E}(\rho)S_i+S_i\mathcal{E}(\rho)}{2}\delta].
\end{eqnarray}
Substituting Eq.~(\ref{SLD}) into Eq.~(\ref{exchange-order}) and by definition, we have
\begin{eqnarray}\label[step]{step1}
    \langle S_i,\delta \rangle_{\mathcal{E}(\rho)}=\tr[\left(\frac{\partial}{\partial\theta_i}\mathcal{E}(\rho)\right)\delta].
\end{eqnarray}
Inserting the expression of $\delta$ into Eq.~(\ref{step1}), we obtain that
\begin{eqnarray}\label[step]{step2}
    \langle S_i,\delta \rangle_{\mathcal{E}(\rho)}=\tr[\left(\frac{\partial}{\partial\theta_i}\mathcal{E}(\rho)\right)\mathcal{E}^{*-1}(X)],
\end{eqnarray}
where we have used the equality $\tr[\frac{\partial}{\partial\theta_i}\mathcal{E}(\rho)]=0$. Noting that
$\tr[\left(\frac{\partial}{\partial\theta_i}\mathcal{E}(\rho)\right)\mathcal{E}^{*-1}(X)]=\frac{\partial}{\partial\theta_i}\tr[\mathcal{E}(\rho)\mathcal{E}^{*-1}(X)]=\frac{\partial}{\partial\theta_i}\tr(\rho X)=\frac{\partial}{\partial\theta_i}\overline{X}$, we deduce from Eq.~(\ref{step2}) that the $\delta$ in Eq.~(\ref{eq_construct_delta}) satisfies Eq.~(\ref{eq_delta_locallyunbiased_condition}).

\section{Proof of $\mathcal{P}(h)=0$} \label{appsec_proofPhZero}

Since $\mathcal{P}(h)$ commutes with all $U_g$, we can express $\mathcal{P}(h)$ as
\begin{equation} \label{appeq_parameterize_Ph}
    \mathcal{P} \left( h \right) =\bigoplus_{\alpha =1}^s{\left( a_{\alpha}\mathbb{1} _{n_{\alpha}}+\boldsymbol{b}_{\alpha}\cdot \boldsymbol{\lambda }_{\alpha} \right)}\otimes \mathbb{1} _{d_{\alpha}},
\end{equation}
where $a_{\alpha}$ and $\boldsymbol{b}_{\alpha}=\left( b_{\alpha ,1},\cdots ,b_{\alpha ,n_{\alpha}^{2}-1} \right) $,  $\alpha=1,...,s$, are some real parameters to be determined. Note that $\mathcal{P}(h)$ satisfies
\begin{eqnarray}\label{App-1}
    \tr\left[\frac{\partial\rho}{\partial\theta_i}\mathcal{P}(h)\right]=0.
\end{eqnarray}
Inserting \cref{appeq_parameterization_rho,appeq_theta_parameterize,appeq_parameterize_Ph} into Eq.~(\ref{App-1}) and utilizing the algebraic properties of $\lambda_{\alpha,i}$ [see \cref{appeq_generator_relation}], we have
\begin{equation}\label{app-2}
    \mathrm{tr}\left[ \mathcal{P} \left( h \right) \left( \partial \rho /\partial r_{\alpha ,i} \right) \right] =b_{\alpha ,i}=0
\end{equation}
for $\alpha=1,...,s$ and $i=1,...,n_{\alpha}^{2}-1$, and
\begin{equation}\label{app-3}
    \mathrm{tr}\left[ \mathcal{P} \left( h \right) \left( \partial \rho /\partial p_{\alpha} \right) \right] =a_{\alpha}-a_s=0
\end{equation}
for $\alpha=1,...,s-1$. From \cref{app-2,app-3}, it follows that $\mathcal{P} \left( h \right)$ is proportional to the identity matrix. Lastly, noting that $\mathcal{P}(h)$ satisfies $\mathrm{tr}\left[ \rho \mathcal{P} \left( h \right) \right] =0$, we have $\mathcal{P}(h)=0$.

\section{Canonical SDP form of \cref{eq_sdpform}} \label{appsec_proof_eq_sdpform_SDP}
Let $\Phi$ be a Hermitian-preserving map and $\mathsf{A},\mathsf{B}$ be Hermitian operators. An SDP is a triple $(\Phi,\mathsf{A},\mathsf{B})$ with which the following optimization problem is associated \cite{Wat18}
\begin{equation} \label{appeq_sdp_canonical}
    \begin{aligned}
        \underset{\mathsf{Y}}{\min}\quad & \mathrm{tr}\left( \mathsf{BY} \right)
        \\
        \mathrm{s}.\mathrm{t}.\quad      & \Phi \left( \mathsf{Y} \right) \ge \mathsf{A},
        \\
                                            & \mathsf{Y}=\mathsf{Y}^{\dagger}.
    \end{aligned}
\end{equation}
Here, to distinguish the symbol used in Eq.~(\ref{matrix-A}), we have adopted the symbol $\mathsf{A}$. To reformulate \cref{eq_sdpform} into the canonical SDP form given by \cref{appeq_sdp_canonical}, we partition $\mathsf{Y}$ as
\begin{equation}
    \mathsf{Y}=\left[ \begin{matrix}
            \mathsf{Y}_{11} & \mathsf{Y}_{12} \\
            \mathsf{Y}_{21} & \mathsf{Y}_{22} \\
        \end{matrix} \right] ,
\end{equation}
where $\mathsf{Y}_{11}=\Lambda$ and $\mathsf{Y}_{22}=h$, and $\mathsf{Y}_{12}$ and $\mathsf{Y}_{21}$ are dummy variables. We introduce the Hermitian-preserving map
\begin{equation}
    \Phi \left( \mathsf{Y} \right) =\left[ \begin{matrix}
            \mathsf{Y}_{11}                                                              & -\mathcal{E} ^{*-1}\left( \mathcal{Q} \left( \mathsf{Y}_{22} \right) \right) \\
            -\mathcal{E} ^{*-1}\left( \mathcal{Q} \left( \mathsf{Y}_{22} \right) \right) & 0                                                                            \\
        \end{matrix} \right] ,
\end{equation}
and specify $\mathsf{A}$ and $\mathsf{B}$ to be
\begin{equation}
    \mathsf{A}=\left[ \begin{matrix}
            0                                                          & \overline{X}\mathbb{1} -\mathcal{E} ^{*-1}\left( X \right) \\
            \overline{X}\mathbb{1} -\mathcal{E} ^{*-1}\left( X \right) & -\mathbb{1}                                                \\
        \end{matrix} \right] ,\quad \mathsf{B}=\left[ \begin{matrix}
            \mathcal{E} \left( \rho \right) & 0 \\
            0                               & 0 \\
        \end{matrix} \right] .
\end{equation}
Using the above $\Phi,\mathsf{A}$ and $\mathsf{B}$, we can straightforwardly verify the equivalence between \cref{eq_sdpform} and the SDP in \cref{appeq_sdp_canonical}.

\section{Intuitive understanding of \cref{th_covariant}} \label{appsec_intuitive}
\cref{th_covariant} may be understood as follows. $\overline X$ can be reformulated as
\begin{equation}
    \overline{X}=\mathrm{tr}\left[ \mathcal{E} \left( \rho \right) \mathcal{E} ^{*-1}\left( X \right) \right],
\end{equation}
which can be understood as the expectation value of the observable $\mathcal{E} ^{*-1}\left( X \right)$ in $\mathcal{E} \left( \rho \right) $. Besides, when $\mathcal{E}$ is covariant, 
\begin{equation}
    U_g\mathcal{E} \left( \rho \right) U_{g}^{\dagger}=\mathcal{E} \left( U_g\rho U_{g}^{\dagger} \right) =\mathcal{E} \left( \rho \right) ,
\end{equation}
for all $g\in G$. This implies that $\mathcal{E}(\rho)$ is with the symmetric structures described by $G$. Then, the result from Ref.~\cite{ZT24} [i.e.~Eq.~(2) therein] suggests that $Y_{h_0}=\mathcal{P} \left( \mathcal{E} ^{*-1}\left( X \right) \right)$.


\begin{thebibliography}{70}%
    \makeatletter
    \providecommand \@ifxundefined [1]{%
        \@ifx{#1\undefined}
    }%
    \providecommand \@ifnum [1]{%
        \ifnum #1\expandafter \@firstoftwo
        \else \expandafter \@secondoftwo
        \fi
    }%
    \providecommand \@ifx [1]{%
        \ifx #1\expandafter \@firstoftwo
        \else \expandafter \@secondoftwo
        \fi
    }%
    \providecommand \natexlab [1]{#1}%
    \providecommand \enquote  [1]{``#1''}%
    \providecommand \bibnamefont  [1]{#1}%
    \providecommand \bibfnamefont [1]{#1}%
    \providecommand \citenamefont [1]{#1}%
    \providecommand \href@noop [0]{\@secondoftwo}%
    \providecommand \href [0]{\begingroup \@sanitize@url \@href}%
    \providecommand \@href[1]{\@@startlink{#1}\@@href}%
    \providecommand \@@href[1]{\endgroup#1\@@endlink}%
    \providecommand \@sanitize@url [0]{\catcode `\\12\catcode `\$12\catcode
        `\&12\catcode `\#12\catcode `\^12\catcode `\_12\catcode `\%12\relax}%
    \providecommand \@@startlink[1]{}%
    \providecommand \@@endlink[0]{}%
    \providecommand \url  [0]{\begingroup\@sanitize@url \@url }%
    \providecommand \@url [1]{\endgroup\@href {#1}{\urlprefix }}%
    \providecommand \urlprefix  [0]{URL }%
    \providecommand \Eprint [0]{\href }%
    \providecommand \doibase [0]{https://doi.org/}%
    \providecommand \selectlanguage [0]{\@gobble}%
    \providecommand \bibinfo  [0]{\@secondoftwo}%
    \providecommand \bibfield  [0]{\@secondoftwo}%
    \providecommand \translation [1]{[#1]}%
    \providecommand \BibitemOpen [0]{}%
    \providecommand \bibitemStop [0]{}%
    \providecommand \bibitemNoStop [0]{.\EOS\space}%
    \providecommand \EOS [0]{\spacefactor3000\relax}%
    \providecommand \BibitemShut  [1]{\csname bibitem#1\endcsname}%
    \let\auto@bib@innerbib\@empty
    \bibitem [{\citenamefont {Gross}\ \emph {et~al.}(2010)\citenamefont {Gross},
                \citenamefont {Liu}, \citenamefont {Flammia}, \citenamefont {Becker},\ and\
                \citenamefont {Eisert}}]{GLF10}%
    \BibitemOpen
    \bibfield  {author} {\bibinfo {author} {\bibfnamefont {D.}~\bibnamefont
            {Gross}}, \bibinfo {author} {\bibfnamefont {Y.-K.}\ \bibnamefont {Liu}},
        \bibinfo {author} {\bibfnamefont {S.~T.}\ \bibnamefont {Flammia}}, \bibinfo
        {author} {\bibfnamefont {S.}~\bibnamefont {Becker}},\ and\ \bibinfo {author}
        {\bibfnamefont {J.}~\bibnamefont {Eisert}},\ }\bibfield  {title} {\bibinfo
        {title} {{Quantum State Tomography via Compressed Sensing}},\ }\href
    {https://doi.org/10.1103/physrevlett.105.150401} {\bibfield  {journal}
        {\bibinfo  {journal} {Phys. Rev. Lett.}\ }\textbf {\bibinfo {volume} {105}},\
        \bibinfo {pages} {150401} (\bibinfo {year} {2010})}\BibitemShut {NoStop}%
    \bibitem [{\citenamefont {Liu}\ \emph {et~al.}(2012)\citenamefont {Liu},
                \citenamefont {Zhang}, \citenamefont {Liu}, \citenamefont {Chen},\ and\
                \citenamefont {Yuan}}]{LZL12}%
    \BibitemOpen
    \bibfield  {author} {\bibinfo {author} {\bibfnamefont {W.-T.}\ \bibnamefont
            {Liu}}, \bibinfo {author} {\bibfnamefont {T.}~\bibnamefont {Zhang}}, \bibinfo
        {author} {\bibfnamefont {J.-Y.}\ \bibnamefont {Liu}}, \bibinfo {author}
        {\bibfnamefont {P.-X.}\ \bibnamefont {Chen}},\ and\ \bibinfo {author}
        {\bibfnamefont {J.-M.}\ \bibnamefont {Yuan}},\ }\bibfield  {title} {\bibinfo
        {title} {{Experimental Quantum State Tomography via Compressed Sampling}},\
    }\href {https://doi.org/10.1103/physrevlett.108.170403} {\bibfield  {journal}
        {\bibinfo  {journal} {Phys. Rev. Lett.}\ }\textbf {\bibinfo {volume} {108}},\
        \bibinfo {pages} {170403} (\bibinfo {year} {2012})}\BibitemShut {NoStop}%
    \bibitem [{\citenamefont {Mahler}\ \emph {et~al.}(2013)\citenamefont {Mahler},
                \citenamefont {Rozema}, \citenamefont {Darabi}, \citenamefont {Ferrie},
                \citenamefont {{Blume-Kohout}},\ and\ \citenamefont
                {Steinberg}}]{Mahler2013PRL}%
    \BibitemOpen
    \bibfield  {author} {\bibinfo {author} {\bibfnamefont {D.~H.}\ \bibnamefont
            {Mahler}}, \bibinfo {author} {\bibfnamefont {L.~A.}\ \bibnamefont {Rozema}},
        \bibinfo {author} {\bibfnamefont {A.}~\bibnamefont {Darabi}}, \bibinfo
        {author} {\bibfnamefont {C.}~\bibnamefont {Ferrie}}, \bibinfo {author}
        {\bibfnamefont {R.}~\bibnamefont {{Blume-Kohout}}},\ and\ \bibinfo {author}
        {\bibfnamefont {A.~M.}\ \bibnamefont {Steinberg}},\ }\bibfield  {title}
    {\bibinfo {title} {{Adaptive Quantum State Tomography Improves Accuracy
                    Quadratically}},\ }\href {https://doi.org/10.1103/PhysRevLett.111.183601}
    {\bibfield  {journal} {\bibinfo  {journal} {Phys. Rev. Lett.}\ }\textbf
        {\bibinfo {volume} {111}},\ \bibinfo {pages} {183601} (\bibinfo {year}
        {2013})}\BibitemShut {NoStop}%
    \bibitem [{\citenamefont {Qi}\ \emph {et~al.}(2017)\citenamefont {Qi},
                \citenamefont {Hou}, \citenamefont {Wang}, \citenamefont {Dong},
                \citenamefont {Zhong}, \citenamefont {Li}, \citenamefont {Xiang},
                \citenamefont {Wiseman}, \citenamefont {Li},\ and\ \citenamefont
                {Guo}}]{Qi2017nQI}%
    \BibitemOpen
    \bibfield  {author} {\bibinfo {author} {\bibfnamefont {B.}~\bibnamefont
            {Qi}}, \bibinfo {author} {\bibfnamefont {Z.}~\bibnamefont {Hou}}, \bibinfo
        {author} {\bibfnamefont {Y.}~\bibnamefont {Wang}}, \bibinfo {author}
        {\bibfnamefont {D.}~\bibnamefont {Dong}}, \bibinfo {author} {\bibfnamefont
            {H.-S.}\ \bibnamefont {Zhong}}, \bibinfo {author} {\bibfnamefont
            {L.}~\bibnamefont {Li}}, \bibinfo {author} {\bibfnamefont {G.-Y.}\
            \bibnamefont {Xiang}}, \bibinfo {author} {\bibfnamefont {H.~M.}\ \bibnamefont
            {Wiseman}}, \bibinfo {author} {\bibfnamefont {C.-F.}\ \bibnamefont {Li}},\
        and\ \bibinfo {author} {\bibfnamefont {G.-C.}\ \bibnamefont {Guo}},\
    }\bibfield  {title} {\bibinfo {title} {{Adaptive quantum state tomography via
                    linear regression estimation: Theory and two-qubit experiment}},\ }\href
    {https://doi.org/10.1038/s41534-017-0016-4} {\bibfield  {journal} {\bibinfo
            {journal} {npj Quantum Inf.}\ }\textbf {\bibinfo {volume} {3}},\ \bibinfo
        {pages} {19} (\bibinfo {year} {2017})}\BibitemShut {NoStop}%
    \bibitem [{\citenamefont {Ferrie}(2014)}]{Ferrie2014PRL}%
    \BibitemOpen
    \bibfield  {author} {\bibinfo {author} {\bibfnamefont {C.}~\bibnamefont
            {Ferrie}},\ }\bibfield  {title} {\bibinfo {title} {{Self-Guided Quantum
                    Tomography}},\ }\href {https://doi.org/10.1103/PhysRevLett.113.190404}
    {\bibfield  {journal} {\bibinfo  {journal} {Phys. Rev. Lett.}\ }\textbf
        {\bibinfo {volume} {113}},\ \bibinfo {pages} {190404} (\bibinfo {year}
        {2014})}\BibitemShut {NoStop}%
    \bibitem [{\citenamefont {Rambach}\ \emph {et~al.}(2021)\citenamefont
                {Rambach}, \citenamefont {Qaryan}, \citenamefont {Kewming}, \citenamefont
                {Ferrie}, \citenamefont {White},\ and\ \citenamefont
                {Romero}}]{Rambach2021PRL}%
    \BibitemOpen
    \bibfield  {author} {\bibinfo {author} {\bibfnamefont {M.}~\bibnamefont
            {Rambach}}, \bibinfo {author} {\bibfnamefont {M.}~\bibnamefont {Qaryan}},
        \bibinfo {author} {\bibfnamefont {M.}~\bibnamefont {Kewming}}, \bibinfo
        {author} {\bibfnamefont {C.}~\bibnamefont {Ferrie}}, \bibinfo {author}
        {\bibfnamefont {A.~G.}\ \bibnamefont {White}},\ and\ \bibinfo {author}
        {\bibfnamefont {J.}~\bibnamefont {Romero}},\ }\bibfield  {title} {\bibinfo
        {title} {{Robust and Efficient High-Dimensional Quantum State Tomography}},\
    }\href {https://doi.org/10.1103/PhysRevLett.126.100402} {\bibfield  {journal}
        {\bibinfo  {journal} {Phys. Rev. Lett.}\ }\textbf {\bibinfo {volume} {126}},\
        \bibinfo {pages} {100402} (\bibinfo {year} {2021})}\BibitemShut {NoStop}%
    \bibitem [{\citenamefont {Aaronson}(2018)}]{Aaronson2018}%
    \BibitemOpen
    \bibfield  {author} {\bibinfo {author} {\bibfnamefont {S.}~\bibnamefont
            {Aaronson}},\ }\bibfield  {title} {\bibinfo {title} {Shadow tomography of
            quantum states},\ }in\ \href {https://doi.org/10.1145/3188745.3188802} {\emph
        {\bibinfo {booktitle} {Proceedings of the 50th Annual {ACM} {SIGACT}
                Symposium on Theory of Computing}}}\ (\bibinfo  {publisher} {{ACM, New
                York}},\ \bibinfo {year} {2018})\ pp.\ \bibinfo {pages}
    {325--338}\BibitemShut {NoStop}%
    \bibitem [{\citenamefont {Huang}\ \emph {et~al.}(2020)\citenamefont {Huang},
                \citenamefont {Kueng},\ and\ \citenamefont {Preskill}}]{HKP20}%
    \BibitemOpen
    \bibfield  {author} {\bibinfo {author} {\bibfnamefont {H.-Y.}\ \bibnamefont
            {Huang}}, \bibinfo {author} {\bibfnamefont {R.}~\bibnamefont {Kueng}},\ and\
        \bibinfo {author} {\bibfnamefont {J.}~\bibnamefont {Preskill}},\ }\bibfield
    {title} {\bibinfo {title} {Predicting many properties of a quantum system
            from very few measurements},\ }\href
    {https://doi.org/10.1038/s41567-020-0932-7} {\bibfield  {journal} {\bibinfo
            {journal} {Nat. Phys.}\ }\textbf {\bibinfo {volume} {16}},\ \bibinfo {pages}
        {1050} (\bibinfo {year} {2020})}\BibitemShut {NoStop}%
    \bibitem [{\citenamefont {Huang}\ \emph {et~al.}(2021)\citenamefont {Huang},
                \citenamefont {Kueng},\ and\ \citenamefont {Preskill}}]{HKP21}%
    \BibitemOpen
    \bibfield  {author} {\bibinfo {author} {\bibfnamefont {H.-Y.}\ \bibnamefont
            {Huang}}, \bibinfo {author} {\bibfnamefont {R.}~\bibnamefont {Kueng}},\ and\
        \bibinfo {author} {\bibfnamefont {J.}~\bibnamefont {Preskill}},\ }\bibfield
    {title} {\bibinfo {title} {{Efficient Estimation of Pauli Observables by
                    Derandomization}},\ }\href {https://doi.org/10.1103/physrevlett.127.030503}
    {\bibfield  {journal} {\bibinfo  {journal} {Phys. Rev. Lett.}\ }\textbf
        {\bibinfo {volume} {127}},\ \bibinfo {pages} {030503} (\bibinfo {year}
        {2021})}\BibitemShut {NoStop}%
    \bibitem [{\citenamefont {Zhang}\ and\ \citenamefont {Tong}(2024)}]{ZT24}%
    \BibitemOpen
    \bibfield  {author} {\bibinfo {author} {\bibfnamefont {D.-J.}\ \bibnamefont
            {Zhang}}\ and\ \bibinfo {author} {\bibfnamefont {D.~M.}\ \bibnamefont
            {Tong}},\ }\bibfield  {title} {\bibinfo {title} {{Inferring Physical
                    Properties of Symmetric States from the Fewest Copies}},\ }\href
    {https://doi.org/10.1103/physrevlett.133.040202} {\bibfield  {journal}
        {\bibinfo  {journal} {Phys. Rev. Lett.}\ }\textbf {\bibinfo {volume} {133}},\
        \bibinfo {pages} {040202} (\bibinfo {year} {2024})}\BibitemShut {NoStop}%
    \bibitem [{\citenamefont {Zhou}\ and\ \citenamefont {Zhang}(2023)}]{Zhou2023E}%
    \BibitemOpen
    \bibfield  {author} {\bibinfo {author} {\bibfnamefont {L.}~\bibnamefont
            {Zhou}}\ and\ \bibinfo {author} {\bibfnamefont {D.-J.}\ \bibnamefont
            {Zhang}},\ }\bibfield  {title} {\bibinfo {title} {{Non-{Hermitian} {Floquet}
    Topological Matter—A Review}},\ }\href {https://doi.org/10.3390/e25101401}
    {\bibfield  {journal} {\bibinfo  {journal} {Entropy}\ }\textbf {\bibinfo
            {volume} {25}},\ \bibinfo {pages} {1401} (\bibinfo {year}
        {2023})}\BibitemShut {NoStop}%
    \bibitem [{\citenamefont {Kiesel}\ \emph {et~al.}(2007)\citenamefont {Kiesel},
                \citenamefont {Schmid}, \citenamefont {T{\'o}th}, \citenamefont {Solano},\
                and\ \citenamefont {Weinfurter}}]{KST07}%
    \BibitemOpen
    \bibfield  {author} {\bibinfo {author} {\bibfnamefont {N.}~\bibnamefont
            {Kiesel}}, \bibinfo {author} {\bibfnamefont {C.}~\bibnamefont {Schmid}},
        \bibinfo {author} {\bibfnamefont {G.}~\bibnamefont {T{\'o}th}}, \bibinfo
        {author} {\bibfnamefont {E.}~\bibnamefont {Solano}},\ and\ \bibinfo {author}
        {\bibfnamefont {H.}~\bibnamefont {Weinfurter}},\ }\bibfield  {title}
    {\bibinfo {title} {{Experimental Observation of Four-Photon Entangled {D}icke
                    State with High Fidelity}},\ }\href
    {https://doi.org/10.1103/physrevlett.98.063604} {\bibfield  {journal}
        {\bibinfo  {journal} {Phys. Rev. Lett.}\ }\textbf {\bibinfo {volume} {98}},\
        \bibinfo {pages} {063604} (\bibinfo {year} {2007})}\BibitemShut {NoStop}%
    \bibitem [{\citenamefont {Wieczorek}\ \emph {et~al.}(2008)\citenamefont
                {Wieczorek}, \citenamefont {Schmid}, \citenamefont {Kiesel}, \citenamefont
                {Pohlner}, \citenamefont {G{\"u}hne},\ and\ \citenamefont
                {Weinfurter}}]{WSK08}%
    \BibitemOpen
    \bibfield  {author} {\bibinfo {author} {\bibfnamefont {W.}~\bibnamefont
            {Wieczorek}}, \bibinfo {author} {\bibfnamefont {C.}~\bibnamefont {Schmid}},
        \bibinfo {author} {\bibfnamefont {N.}~\bibnamefont {Kiesel}}, \bibinfo
        {author} {\bibfnamefont {R.}~\bibnamefont {Pohlner}}, \bibinfo {author}
        {\bibfnamefont {O.}~\bibnamefont {G{\"u}hne}},\ and\ \bibinfo {author}
        {\bibfnamefont {H.}~\bibnamefont {Weinfurter}},\ }\bibfield  {title}
    {\bibinfo {title} {{Experimental Observation of an Entire Family of
                    Four-Photon Entangled States}},\ }\href
    {https://doi.org/10.1103/physrevlett.101.010503} {\bibfield  {journal}
        {\bibinfo  {journal} {Phys. Rev. Lett.}\ }\textbf {\bibinfo {volume} {101}},\
        \bibinfo {pages} {010503} (\bibinfo {year} {2008})}\BibitemShut {NoStop}%
    \bibitem [{\citenamefont {Wieczorek}\ \emph {et~al.}(2009)\citenamefont
                {Wieczorek}, \citenamefont {Krischek}, \citenamefont {Kiesel}, \citenamefont
                {Michelberger}, \citenamefont {T{\'o}th},\ and\ \citenamefont
                {Weinfurter}}]{WKK09}%
    \BibitemOpen
    \bibfield  {author} {\bibinfo {author} {\bibfnamefont {W.}~\bibnamefont
            {Wieczorek}}, \bibinfo {author} {\bibfnamefont {R.}~\bibnamefont {Krischek}},
        \bibinfo {author} {\bibfnamefont {N.}~\bibnamefont {Kiesel}}, \bibinfo
        {author} {\bibfnamefont {P.}~\bibnamefont {Michelberger}}, \bibinfo {author}
        {\bibfnamefont {G.}~\bibnamefont {T{\'o}th}},\ and\ \bibinfo {author}
        {\bibfnamefont {H.}~\bibnamefont {Weinfurter}},\ }\bibfield  {title}
    {\bibinfo {title} {{Experimental Entanglement of a Six-Photon Symmetric Dicke
                    State}},\ }\href {https://doi.org/10.1103/physrevlett.103.020504} {\bibfield
        {journal} {\bibinfo  {journal} {Phys. Rev. Lett.}\ }\textbf {\bibinfo
            {volume} {103}},\ \bibinfo {pages} {020504} (\bibinfo {year}
        {2009})}\BibitemShut {NoStop}%
    \bibitem [{\citenamefont {Prevedel}\ \emph {et~al.}(2009)\citenamefont
                {Prevedel}, \citenamefont {Cronenberg}, \citenamefont {Tame}, \citenamefont
                {Paternostro}, \citenamefont {Walther}, \citenamefont {Kim},\ and\
                \citenamefont {Zeilinger}}]{PCT09}%
    \BibitemOpen
    \bibfield  {author} {\bibinfo {author} {\bibfnamefont {R.}~\bibnamefont
            {Prevedel}}, \bibinfo {author} {\bibfnamefont {G.}~\bibnamefont
            {Cronenberg}}, \bibinfo {author} {\bibfnamefont {M.~S.}\ \bibnamefont
            {Tame}}, \bibinfo {author} {\bibfnamefont {M.}~\bibnamefont {Paternostro}},
        \bibinfo {author} {\bibfnamefont {P.}~\bibnamefont {Walther}}, \bibinfo
        {author} {\bibfnamefont {M.~S.}\ \bibnamefont {Kim}},\ and\ \bibinfo {author}
        {\bibfnamefont {A.}~\bibnamefont {Zeilinger}},\ }\bibfield  {title} {\bibinfo
        {title} {{Experimental Realization of Dicke States of up to Six Qubits for
                    Multiparty Quantum Networking}},\ }\href
    {https://doi.org/10.1103/physrevlett.103.020503} {\bibfield  {journal}
        {\bibinfo  {journal} {Phys. Rev. Lett.}\ }\textbf {\bibinfo {volume} {103}},\
        \bibinfo {pages} {020503} (\bibinfo {year} {2009})}\BibitemShut {NoStop}%
    \bibitem [{\citenamefont {Krischek}\ \emph {et~al.}(2010)\citenamefont
                {Krischek}, \citenamefont {Wieczorek}, \citenamefont {Ozawa}, \citenamefont
                {Kiesel}, \citenamefont {Michelberger}, \citenamefont {Udem},\ and\
                \citenamefont {Weinfurter}}]{KWO10}%
    \BibitemOpen
    \bibfield  {author} {\bibinfo {author} {\bibfnamefont {R.}~\bibnamefont
            {Krischek}}, \bibinfo {author} {\bibfnamefont {W.}~\bibnamefont {Wieczorek}},
        \bibinfo {author} {\bibfnamefont {A.}~\bibnamefont {Ozawa}}, \bibinfo
        {author} {\bibfnamefont {N.}~\bibnamefont {Kiesel}}, \bibinfo {author}
        {\bibfnamefont {P.}~\bibnamefont {Michelberger}}, \bibinfo {author}
        {\bibfnamefont {T.}~\bibnamefont {Udem}},\ and\ \bibinfo {author}
        {\bibfnamefont {H.}~\bibnamefont {Weinfurter}},\ }\bibfield  {title}
    {\bibinfo {title} {Ultraviolet enhancement cavity for ultrafast nonlinear
            optics and high-rate multiphoton entanglement experiments},\ }\href
    {https://doi.org/10.1038/nphoton.2009.286} {\bibfield  {journal} {\bibinfo
            {journal} {Nature Photon.}\ }\textbf {\bibinfo {volume} {4}},\ \bibinfo
        {pages} {170} (\bibinfo {year} {2010})}\BibitemShut {NoStop}%
    \bibitem [{\citenamefont {Erhard}\ \emph {et~al.}(2018)\citenamefont {Erhard},
                \citenamefont {Malik}, \citenamefont {Krenn},\ and\ \citenamefont
                {Zeilinger}}]{EMKZ18}%
    \BibitemOpen
    \bibfield  {author} {\bibinfo {author} {\bibfnamefont {M.}~\bibnamefont
            {Erhard}}, \bibinfo {author} {\bibfnamefont {M.}~\bibnamefont {Malik}},
        \bibinfo {author} {\bibfnamefont {M.}~\bibnamefont {Krenn}},\ and\ \bibinfo
        {author} {\bibfnamefont {A.}~\bibnamefont {Zeilinger}},\ }\bibfield  {title}
    {\bibinfo {title} {{{E}xperimental {G}reenberger–{H}orne–{Z}eilinger
    entanglement beyond qubits}},\ }\href
    {https://doi.org/10.1038/s41566-018-0257-6} {\bibfield  {journal} {\bibinfo
            {journal} {Nature Photon.}\ }\textbf {\bibinfo {volume} {12}},\ \bibinfo
        {pages} {759} (\bibinfo {year} {2018})}\BibitemShut {NoStop}%
    \bibitem [{\citenamefont {Liu}\ \emph {et~al.}(2021)\citenamefont {Liu},
                \citenamefont {Zhou}, \citenamefont {Meng}, \citenamefont {Yang},
                \citenamefont {Li}, \citenamefont {Meng}, \citenamefont {Su}, \citenamefont
                {Chen}, \citenamefont {Sun}, \citenamefont {Xu}, \citenamefont {Li},\ and\
                \citenamefont {Guo}}]{LZM21}%
    \BibitemOpen
    \bibfield  {author} {\bibinfo {author} {\bibfnamefont {Z.-H.}\ \bibnamefont
            {Liu}}, \bibinfo {author} {\bibfnamefont {J.}~\bibnamefont {Zhou}}, \bibinfo
        {author} {\bibfnamefont {H.-X.}\ \bibnamefont {Meng}}, \bibinfo {author}
        {\bibfnamefont {M.}~\bibnamefont {Yang}}, \bibinfo {author} {\bibfnamefont
            {Q.}~\bibnamefont {Li}}, \bibinfo {author} {\bibfnamefont {Y.}~\bibnamefont
            {Meng}}, \bibinfo {author} {\bibfnamefont {H.-Y.}\ \bibnamefont {Su}},
        \bibinfo {author} {\bibfnamefont {J.-L.}\ \bibnamefont {Chen}}, \bibinfo
        {author} {\bibfnamefont {K.}~\bibnamefont {Sun}}, \bibinfo {author}
        {\bibfnamefont {J.-S.}\ \bibnamefont {Xu}}, \bibinfo {author} {\bibfnamefont
            {C.-F.}\ \bibnamefont {Li}},\ and\ \bibinfo {author} {\bibfnamefont {G.-C.}\
            \bibnamefont {Guo}},\ }\bibfield  {title} {\bibinfo {title} {{E}xperimental
    test of the {G}reenberger–{H}orne–{Z}eilinger-type paradoxes in and
    beyond graph states},\ }\href {https://doi.org/10.1038/s41534-021-00397-z}
    {\bibfield  {journal} {\bibinfo  {journal} {npj Quantum Inf.}\ }\textbf
        {\bibinfo {volume} {7}},\ \bibinfo {pages} {66} (\bibinfo {year}
        {2021})}\BibitemShut {NoStop}%
    \bibitem [{\citenamefont {T{\'o}th}\ \emph {et~al.}(2010)\citenamefont
                {T{\'o}th}, \citenamefont {Wieczorek}, \citenamefont {Gross}, \citenamefont
                {Krischek}, \citenamefont {Schwemmer},\ and\ \citenamefont
                {Weinfurter}}]{TWG10}%
    \BibitemOpen
    \bibfield  {author} {\bibinfo {author} {\bibfnamefont {G.}~\bibnamefont
            {T{\'o}th}}, \bibinfo {author} {\bibfnamefont {W.}~\bibnamefont {Wieczorek}},
        \bibinfo {author} {\bibfnamefont {D.}~\bibnamefont {Gross}}, \bibinfo
        {author} {\bibfnamefont {R.}~\bibnamefont {Krischek}}, \bibinfo {author}
        {\bibfnamefont {C.}~\bibnamefont {Schwemmer}},\ and\ \bibinfo {author}
        {\bibfnamefont {H.}~\bibnamefont {Weinfurter}},\ }\bibfield  {title}
    {\bibinfo {title} {{Permutationally Invariant Quantum Tomography}},\ }\href
    {https://doi.org/10.1103/physrevlett.105.250403} {\bibfield  {journal}
        {\bibinfo  {journal} {Phys. Rev. Lett.}\ }\textbf {\bibinfo {volume} {105}},\
        \bibinfo {pages} {250403} (\bibinfo {year} {2010})}\BibitemShut {NoStop}%
    \bibitem [{\citenamefont {Moroder}\ \emph {et~al.}(2012)\citenamefont
                {Moroder}, \citenamefont {Hyllus}, \citenamefont {T{\'o}th}, \citenamefont
                {Schwemmer}, \citenamefont {Niggebaum}, \citenamefont {Gaile}, \citenamefont
                {G{\"u}hne},\ and\ \citenamefont {Weinfurter}}]{MHT12}%
    \BibitemOpen
    \bibfield  {author} {\bibinfo {author} {\bibfnamefont {T.}~\bibnamefont
            {Moroder}}, \bibinfo {author} {\bibfnamefont {P.}~\bibnamefont {Hyllus}},
        \bibinfo {author} {\bibfnamefont {G.}~\bibnamefont {T{\'o}th}}, \bibinfo
        {author} {\bibfnamefont {C.}~\bibnamefont {Schwemmer}}, \bibinfo {author}
        {\bibfnamefont {A.}~\bibnamefont {Niggebaum}}, \bibinfo {author}
        {\bibfnamefont {S.}~\bibnamefont {Gaile}}, \bibinfo {author} {\bibfnamefont
            {O.}~\bibnamefont {G{\"u}hne}},\ and\ \bibinfo {author} {\bibfnamefont
            {H.}~\bibnamefont {Weinfurter}},\ }\bibfield  {title} {\bibinfo {title}
        {{Permutationally invariant state reconstruction}},\ }\href
    {https://doi.org/10.1088/1367-2630/14/10/105001} {\bibfield  {journal}
        {\bibinfo  {journal} {New J. Phys.}\ }\textbf {\bibinfo {volume} {14}},\
        \bibinfo {pages} {105001} (\bibinfo {year} {2012})}\BibitemShut {NoStop}%
    \bibitem [{\citenamefont {Gao}\ \emph {et~al.}(2014)\citenamefont {Gao},
                \citenamefont {Yan},\ and\ \citenamefont {van Enk}}]{GYvE14}%
    \BibitemOpen
    \bibfield  {author} {\bibinfo {author} {\bibfnamefont {T.}~\bibnamefont
            {Gao}}, \bibinfo {author} {\bibfnamefont {F.}~\bibnamefont {Yan}},\ and\
        \bibinfo {author} {\bibfnamefont {S.}~\bibnamefont {van Enk}},\ }\bibfield
    {title} {\bibinfo {title} {{Permutationally Invariant Part of a Density
                    Matrix and Nonseparability of {$N$}-Qubit States}},\ }\href
    {https://doi.org/10.1103/physrevlett.112.180501} {\bibfield  {journal}
        {\bibinfo  {journal} {Phys. Rev. Lett.}\ }\textbf {\bibinfo {volume} {112}},\
        \bibinfo {pages} {180501} (\bibinfo {year} {2014})}\BibitemShut {NoStop}%
    \bibitem [{\citenamefont {Werner}(1989)}]{Wer89}%
    \BibitemOpen
    \bibfield  {author} {\bibinfo {author} {\bibfnamefont {R.~F.}\ \bibnamefont
            {Werner}},\ }\bibfield  {title} {\bibinfo {title} {{Quantum states with
        {E}instein-{P}odolsky-{R}osen correlations admitting a hidden-variable
    model}},\ }\href {https://doi.org/10.1103/physreva.40.4277} {\bibfield
        {journal} {\bibinfo  {journal} {Phys. Rev. A}\ }\textbf {\bibinfo {volume}
            {40}},\ \bibinfo {pages} {4277} (\bibinfo {year} {1989})}\BibitemShut
    {NoStop}%
    \bibitem [{\citenamefont {Dicke}(1954)}]{Dic54}%
    \BibitemOpen
    \bibfield  {author} {\bibinfo {author} {\bibfnamefont {R.~H.}\ \bibnamefont
            {Dicke}},\ }\bibfield  {title} {\bibinfo {title} {{Coherence in Spontaneous
                    Radiation Processes}},\ }\href {https://doi.org/10.1103/physrev.93.99}
    {\bibfield  {journal} {\bibinfo  {journal} {Phys. Rev.}\ }\textbf {\bibinfo
            {volume} {93}},\ \bibinfo {pages} {99} (\bibinfo {year} {1954})}\BibitemShut
    {NoStop}%
    \bibitem [{\citenamefont {Greenberger}\ \emph {et~al.}(1989)\citenamefont
                {Greenberger}, \citenamefont {Horne},\ and\ \citenamefont
                {Zeilinger}}]{GHZ89}%
    \BibitemOpen
    \bibfield  {author} {\bibinfo {author} {\bibfnamefont {D.~M.}\ \bibnamefont
            {Greenberger}}, \bibinfo {author} {\bibfnamefont {M.~A.}\ \bibnamefont
            {Horne}},\ and\ \bibinfo {author} {\bibfnamefont {A.}~\bibnamefont
            {Zeilinger}},\ }\href@noop {} {\emph {\bibinfo {title} {{Bell’s Theorem,
                        Quantum Theory, and Conceptions of the Universe}}}},\ edited by\ \bibinfo
    {editor} {\bibfnamefont {M.}~\bibnamefont {Kafatos}}\ (\bibinfo  {publisher}
    {Kluwer Academics, Dordrecht, The Netherlands},\ \bibinfo {year}
    {1989})\BibitemShut {NoStop}%
    \bibitem [{\citenamefont {Horodecki}\ \emph {et~al.}(2009)\citenamefont
                {Horodecki}, \citenamefont {Horodecki}, \citenamefont {Horodecki},\ and\
                \citenamefont {Horodecki}}]{HHHH09}%
    \BibitemOpen
    \bibfield  {author} {\bibinfo {author} {\bibfnamefont {R.}~\bibnamefont
            {Horodecki}}, \bibinfo {author} {\bibfnamefont {P.}~\bibnamefont
            {Horodecki}}, \bibinfo {author} {\bibfnamefont {M.}~\bibnamefont
            {Horodecki}},\ and\ \bibinfo {author} {\bibfnamefont {K.}~\bibnamefont
            {Horodecki}},\ }\bibfield  {title} {\bibinfo {title} {Quantum entanglement},\
    }\href {https://doi.org/10.1103/revmodphys.81.865} {\bibfield  {journal}
        {\bibinfo  {journal} {Rev. Mod. Phys.}\ }\textbf {\bibinfo {volume} {81}},\
        \bibinfo {pages} {865} (\bibinfo {year} {2009})}\BibitemShut {NoStop}%
    \bibitem [{\citenamefont {Gühne}\ and\ \citenamefont
    {T{\'{o}}th}(2009)}]{GT09}%
    \BibitemOpen
    \bibfield  {author} {\bibinfo {author} {\bibfnamefont {O.}~\bibnamefont
        {Gühne}}\ and\ \bibinfo {author} {\bibfnamefont {G.}~\bibnamefont
    {T{\'{o}}th}},\ }\bibfield  {title} {\bibinfo {title} {Entanglement
            detection},\ }\href {https://doi.org/10.1016/j.physrep.2009.02.004}
    {\bibfield  {journal} {\bibinfo  {journal} {Phys. Rep.}\ }\textbf {\bibinfo
            {volume} {474}},\ \bibinfo {pages} {1} (\bibinfo {year} {2009})}\BibitemShut
    {NoStop}%
    \bibitem [{\citenamefont {Hu}\ \emph {et~al.}(2018)\citenamefont {Hu},
                \citenamefont {Hu}, \citenamefont {Wang}, \citenamefont {Peng}, \citenamefont
                {Zhang},\ and\ \citenamefont {Fan}}]{Hu2018PR}%
    \BibitemOpen
    \bibfield  {author} {\bibinfo {author} {\bibfnamefont {M.-L.}\ \bibnamefont
            {Hu}}, \bibinfo {author} {\bibfnamefont {X.}~\bibnamefont {Hu}}, \bibinfo
        {author} {\bibfnamefont {J.}~\bibnamefont {Wang}}, \bibinfo {author}
        {\bibfnamefont {Y.}~\bibnamefont {Peng}}, \bibinfo {author} {\bibfnamefont
            {Y.-R.}\ \bibnamefont {Zhang}},\ and\ \bibinfo {author} {\bibfnamefont
            {H.}~\bibnamefont {Fan}},\ }\bibfield  {title} {\bibinfo {title} {Quantum
            coherence and geometric quantum discord},\ }\href
    {https://doi.org/10.1016/j.physrep.2018.07.004} {\bibfield  {journal}
        {\bibinfo  {journal} {Phys. Rep.}\ }\textbf {\bibinfo {volume} {762}},\
        \bibinfo {pages} {1} (\bibinfo {year} {2018})}\BibitemShut {NoStop}%
    \bibitem [{\citenamefont {Halmos}(1950)}]{Hal50}%
    \BibitemOpen
    \bibfield  {author} {\bibinfo {author} {\bibfnamefont {P.~R.}\ \bibnamefont
            {Halmos}},\ }\href@noop {} {\emph {\bibinfo {title} {Measure Theory}}}\
    (\bibinfo  {publisher} {Springer-Verlag},\ \bibinfo {address} {New York},\
    \bibinfo {year} {1950})\BibitemShut {NoStop}%
    \bibitem [{\citenamefont {Mele}(2024)}]{Mel24}%
    \BibitemOpen
    \bibfield  {author} {\bibinfo {author} {\bibfnamefont {A.~A.}\ \bibnamefont
            {Mele}},\ }\bibfield  {title} {\bibinfo {title} {Introduction to {H}aar
    {M}easure {T}ools in {Q}uantum {I}nformation: {A} {B}eginner's {T}utorial},\
    }\href {https://doi.org/10.22331/q-2024-05-08-1340} {\bibfield  {journal}
        {\bibinfo  {journal} {{Quantum}}\ }\textbf {\bibinfo {volume} {8}},\ \bibinfo
        {pages} {1340} (\bibinfo {year} {2024})}\BibitemShut {NoStop}%
    \bibitem [{\citenamefont {Preskill}(2018)}]{Pre18}%
    \BibitemOpen
    \bibfield  {author} {\bibinfo {author} {\bibfnamefont {J.}~\bibnamefont
            {Preskill}},\ }\bibfield  {title} {\bibinfo {title} {{Quantum Computing in
                    the NISQ era and beyond}},\ }\href {https://doi.org/10.22331/q-2018-08-06-79}
    {\bibfield  {journal} {\bibinfo  {journal} {Quantum}\ }\textbf {\bibinfo
            {volume} {2}},\ \bibinfo {pages} {79} (\bibinfo {year} {2018})}\BibitemShut
    {NoStop}%
    \bibitem [{\citenamefont {Jiao}\ \emph {et~al.}(2023)\citenamefont {Jiao},
                \citenamefont {Wu}, \citenamefont {Bai},\ and\ \citenamefont {An}}]{JWBA23}%
    \BibitemOpen
    \bibfield  {author} {\bibinfo {author} {\bibfnamefont {L.}~\bibnamefont
            {Jiao}}, \bibinfo {author} {\bibfnamefont {W.}~\bibnamefont {Wu}}, \bibinfo
        {author} {\bibfnamefont {S.}~\bibnamefont {Bai}},\ and\ \bibinfo {author}
        {\bibfnamefont {J.}~\bibnamefont {An}},\ }\bibfield  {title} {\bibinfo
        {title} {{Quantum Metrology in the Noisy Intermediate‐Scale Quantum Era}},\
    }\href {https://doi.org/10.1002/qute.202300218} {\bibfield  {journal}
        {\bibinfo  {journal} {Adv. Quantum Technol.}\ }\textbf {\bibinfo {volume}
            {2023}},\ \bibinfo {pages} {2300218} (\bibinfo {year} {2023})}\BibitemShut
    {NoStop}%
    \bibitem [{\citenamefont {Zhang}\ and\ \citenamefont {Tong}(2022)}]{ZT22}%
    \BibitemOpen
    \bibfield  {author} {\bibinfo {author} {\bibfnamefont {D.-J.}\ \bibnamefont
            {Zhang}}\ and\ \bibinfo {author} {\bibfnamefont {D.~M.}\ \bibnamefont
            {Tong}},\ }\bibfield  {title} {\bibinfo {title} {{Approaching
                    Heisenberg-scalable thermometry with built-in robustness against noise}},\
    }\href {https://doi.org/10.1038/s41534-022-00588-2} {\bibfield  {journal}
        {\bibinfo  {journal} {npj Quantum Inf.}\ }\textbf {\bibinfo {volume} {8}},\
        \bibinfo {pages} {81} (\bibinfo {year} {2022})}\BibitemShut {NoStop}%
    \bibitem [{\citenamefont {Ullah}\ \emph {et~al.}(2023)\citenamefont {Ullah},
    \citenamefont {Naseem},\ and\ \citenamefont
    {M\"{u}stecaplio\u{g}lu}}]{UNM23}%
    \BibitemOpen
    \bibfield  {author} {\bibinfo {author} {\bibfnamefont {A.}~\bibnamefont
        {Ullah}}, \bibinfo {author} {\bibfnamefont {M.~T.}\ \bibnamefont {Naseem}},\
    and\ \bibinfo {author} {\bibfnamefont {O.~E.}\ \bibnamefont
    {M\"{u}stecaplio\u{g}lu}},\ }\bibfield  {title} {\bibinfo {title}
        {Low-temperature quantum thermometry boosted by coherence generation},\
    }\href {https://doi.org/10.1103/PhysRevResearch.5.043184} {\bibfield
        {journal} {\bibinfo  {journal} {Phys. Rev. Research}\ }\textbf {\bibinfo
            {volume} {5}},\ \bibinfo {pages} {043184} (\bibinfo {year}
        {2023})}\BibitemShut {NoStop}%
    \bibitem [{\citenamefont {Braunstein}\ and\ \citenamefont
                {Caves}(1994)}]{Braunstein1994PRL}%
    \BibitemOpen
    \bibfield  {author} {\bibinfo {author} {\bibfnamefont {S.~L.}\ \bibnamefont
            {Braunstein}}\ and\ \bibinfo {author} {\bibfnamefont {C.~M.}\ \bibnamefont
            {Caves}},\ }\bibfield  {title} {\bibinfo {title} {{Statistical distance and
                    the geometry of quantum states}},\ }\href
    {https://doi.org/10.1103/PhysRevLett.72.3439} {\bibfield  {journal} {\bibinfo
            {journal} {Phys. Rev. Lett.}\ }\textbf {\bibinfo {volume} {72}},\ \bibinfo
        {pages} {3439} (\bibinfo {year} {1994})}\BibitemShut {NoStop}%
    \bibitem [{\citenamefont {Zhang}\ and\ \citenamefont
                {Gong}(2020)}]{Zhang2020PRR}%
    \BibitemOpen
    \bibfield  {author} {\bibinfo {author} {\bibfnamefont {D.-J.}\ \bibnamefont
            {Zhang}}\ and\ \bibinfo {author} {\bibfnamefont {J.}~\bibnamefont {Gong}},\
    }\bibfield  {title} {\bibinfo {title} {{Dissipative adiabatic measurements:
                    Beating the quantum {Cram\'er-Rao} bound}},\ }\href
    {https://doi.org/10.1103/PhysRevResearch.2.023418} {\bibfield  {journal}
        {\bibinfo  {journal} {Phys. Rev. Research}\ }\textbf {\bibinfo {volume}
            {2}},\ \bibinfo {pages} {023418} (\bibinfo {year} {2020})}\BibitemShut
    {NoStop}%
    \bibitem [{\citenamefont {Tsang}\ \emph {et~al.}(2020)\citenamefont {Tsang},
                \citenamefont {Albarelli},\ and\ \citenamefont {Datta}}]{TAD20}%
    \BibitemOpen
    \bibfield  {author} {\bibinfo {author} {\bibfnamefont {M.}~\bibnamefont
            {Tsang}}, \bibinfo {author} {\bibfnamefont {F.}~\bibnamefont {Albarelli}},\
        and\ \bibinfo {author} {\bibfnamefont {A.}~\bibnamefont {Datta}},\ }\bibfield
    {title} {\bibinfo {title} {{Quantum Semiparametric Estimation}},\ }\href
    {https://doi.org/10.1103/physrevx.10.031023} {\bibfield  {journal} {\bibinfo
            {journal} {Phys. Rev. X}\ }\textbf {\bibinfo {volume} {10}},\ \bibinfo
        {pages} {031023} (\bibinfo {year} {2020})}\BibitemShut {NoStop}%
    \bibitem [{\citenamefont {Zhao}\ \emph {et~al.}(2024)\citenamefont {Zhao},
                \citenamefont {Jing}, \citenamefont {Zhang}, \citenamefont {Zhao},
                \citenamefont {Chen}, \citenamefont {Wang},\ and\ \citenamefont
                {Wang}}]{ZJZ24}%
    \BibitemOpen
    \bibfield  {author} {\bibinfo {author} {\bibfnamefont {B.}~\bibnamefont
            {Zhao}}, \bibinfo {author} {\bibfnamefont {M.}~\bibnamefont {Jing}}, \bibinfo
        {author} {\bibfnamefont {L.}~\bibnamefont {Zhang}}, \bibinfo {author}
        {\bibfnamefont {X.}~\bibnamefont {Zhao}}, \bibinfo {author} {\bibfnamefont
            {Y.-A.}\ \bibnamefont {Chen}}, \bibinfo {author} {\bibfnamefont
            {K.}~\bibnamefont {Wang}},\ and\ \bibinfo {author} {\bibfnamefont
            {X.}~\bibnamefont {Wang}},\ }\bibfield  {title} {\bibinfo {title}
        {{Retrieving Nonlinear Features from Noisy Quantum States}},\ }\href
    {https://doi.org/10.1103/prxquantum.5.020357} {\bibfield  {journal} {\bibinfo
            {journal} {PRX Quantum}\ }\textbf {\bibinfo {volume} {5}},\ \bibinfo {pages}
        {020357} (\bibinfo {year} {2024})}\BibitemShut {NoStop}%
    \bibitem [{\citenamefont {Bartlett}\ \emph {et~al.}(2007)\citenamefont
                {Bartlett}, \citenamefont {Rudolph},\ and\ \citenamefont {Spekkens}}]{BRS07}%
    \BibitemOpen
    \bibfield  {author} {\bibinfo {author} {\bibfnamefont {S.~D.}\ \bibnamefont
            {Bartlett}}, \bibinfo {author} {\bibfnamefont {T.}~\bibnamefont {Rudolph}},\
        and\ \bibinfo {author} {\bibfnamefont {R.~W.}\ \bibnamefont {Spekkens}},\
    }\bibfield  {title} {\bibinfo {title} {{Reference frames, superselection
                    rules, and quantum information}},\ }\href
    {https://doi.org/10.1103/revmodphys.79.555} {\bibfield  {journal} {\bibinfo
            {journal} {Rev. Mod. Phys.}\ }\textbf {\bibinfo {volume} {79}},\ \bibinfo
        {pages} {555} (\bibinfo {year} {2007})}\BibitemShut {NoStop}%
    \bibitem [{\citenamefont {Wick}\ \emph {et~al.}(1952)\citenamefont {Wick},
                \citenamefont {Wightman},\ and\ \citenamefont {Wigner}}]{WWW52}%
    \BibitemOpen
    \bibfield  {author} {\bibinfo {author} {\bibfnamefont {G.~C.}\ \bibnamefont
            {Wick}}, \bibinfo {author} {\bibfnamefont {A.~S.}\ \bibnamefont {Wightman}},\
        and\ \bibinfo {author} {\bibfnamefont {E.~P.}\ \bibnamefont {Wigner}},\
    }\bibfield  {title} {\bibinfo {title} {{The Intrinsic Parity of Elementary
                    Particles}},\ }\href {https://doi.org/10.1103/physrev.88.101} {\bibfield
        {journal} {\bibinfo  {journal} {Phys. Rev.}\ }\textbf {\bibinfo {volume}
            {88}},\ \bibinfo {pages} {101} (\bibinfo {year} {1952})}\BibitemShut
    {NoStop}%
    \bibitem [{\citenamefont {Bartlett}\ and\ \citenamefont
                {Wiseman}(2003)}]{BW03}%
    \BibitemOpen
    \bibfield  {author} {\bibinfo {author} {\bibfnamefont {S.~D.}\ \bibnamefont
            {Bartlett}}\ and\ \bibinfo {author} {\bibfnamefont {H.~M.}\ \bibnamefont
            {Wiseman}},\ }\bibfield  {title} {\bibinfo {title} {{Entanglement Constrained
                    by Superselection Rules}},\ }\href
    {https://doi.org/10.1103/physrevlett.91.097903} {\bibfield  {journal}
        {\bibinfo  {journal} {Phys. Rev. Lett.}\ }\textbf {\bibinfo {volume} {91}},\
        \bibinfo {pages} {097903} (\bibinfo {year} {2003})}\BibitemShut {NoStop}%
    \bibitem [{\citenamefont {Marvian}\ and\ \citenamefont
                {Spekkens}(2014)}]{MS14}%
    \BibitemOpen
    \bibfield  {author} {\bibinfo {author} {\bibfnamefont {I.}~\bibnamefont
            {Marvian}}\ and\ \bibinfo {author} {\bibfnamefont {R.~W.}\ \bibnamefont
            {Spekkens}},\ }\bibfield  {title} {\bibinfo {title} {{Extending Noether’s
                    theorem by quantifying the asymmetry of quantum states}},\ }\href
    {https://doi.org/10.1038/ncomms4821} {\bibfield  {journal} {\bibinfo
            {journal} {Nat. Commun.}\ }\textbf {\bibinfo {volume} {5}},\ \bibinfo {pages}
        {3821} (\bibinfo {year} {2014})}\BibitemShut {NoStop}%
    \bibitem [{\citenamefont {Piani}\ \emph {et~al.}(2016)\citenamefont {Piani},
                \citenamefont {Cianciaruso}, \citenamefont {Bromley}, \citenamefont {Napoli},
                \citenamefont {Johnston},\ and\ \citenamefont {Adesso}}]{PCB16}%
    \BibitemOpen
    \bibfield  {author} {\bibinfo {author} {\bibfnamefont {M.}~\bibnamefont
            {Piani}}, \bibinfo {author} {\bibfnamefont {M.}~\bibnamefont {Cianciaruso}},
        \bibinfo {author} {\bibfnamefont {T.~R.}\ \bibnamefont {Bromley}}, \bibinfo
        {author} {\bibfnamefont {C.}~\bibnamefont {Napoli}}, \bibinfo {author}
        {\bibfnamefont {N.}~\bibnamefont {Johnston}},\ and\ \bibinfo {author}
        {\bibfnamefont {G.}~\bibnamefont {Adesso}},\ }\bibfield  {title} {\bibinfo
        {title} {Robustness of asymmetry and coherence of quantum states},\ }\href
    {https://doi.org/10.1103/PhysRevA.93.042107} {\bibfield  {journal} {\bibinfo
            {journal} {Phys. Rev. A}\ }\textbf {\bibinfo {volume} {93}},\ \bibinfo
        {pages} {042107} (\bibinfo {year} {2016})}\BibitemShut {NoStop}%
    \bibitem [{\citenamefont {Zhang}\ \emph {et~al.}(2017)\citenamefont {Zhang},
                \citenamefont {Yu}, \citenamefont {Huang},\ and\ \citenamefont
                {Tong}}]{ZYHT17}%
    \BibitemOpen
    \bibfield  {author} {\bibinfo {author} {\bibfnamefont {D.-J.}\ \bibnamefont
            {Zhang}}, \bibinfo {author} {\bibfnamefont {X.-D.}\ \bibnamefont {Yu}},
        \bibinfo {author} {\bibfnamefont {H.-L.}\ \bibnamefont {Huang}},\ and\
        \bibinfo {author} {\bibfnamefont {D.~M.}\ \bibnamefont {Tong}},\ }\bibfield
    {title} {\bibinfo {title} {Universal freezing of asymmetry},\ }\href
    {https://doi.org/10.1103/physreva.95.022323} {\bibfield  {journal} {\bibinfo
            {journal} {Phys. Rev. A}\ }\textbf {\bibinfo {volume} {95}},\ \bibinfo
        {pages} {022323} (\bibinfo {year} {2017})}\BibitemShut {NoStop}%
    \bibitem [{\citenamefont {Gour}\ and\ \citenamefont {Spekkens}(2008)}]{GS08}%
    \BibitemOpen
    \bibfield  {author} {\bibinfo {author} {\bibfnamefont {G.}~\bibnamefont
            {Gour}}\ and\ \bibinfo {author} {\bibfnamefont {R.~W.}\ \bibnamefont
            {Spekkens}},\ }\bibfield  {title} {\bibinfo {title} {The resource theory of
            quantum reference frames: manipulations and monotones},\ }\href
    {https://doi.org/10.1088/1367-2630/10/3/033023} {\bibfield  {journal}
        {\bibinfo  {journal} {New J. Phys.}\ }\textbf {\bibinfo {volume} {10}},\
        \bibinfo {pages} {033023} (\bibinfo {year} {2008})}\BibitemShut {NoStop}%
    \bibitem [{\citenamefont {Chitambar}\ and\ \citenamefont {Gour}(2019)}]{CG19}%
    \BibitemOpen
    \bibfield  {author} {\bibinfo {author} {\bibfnamefont {E.}~\bibnamefont
            {Chitambar}}\ and\ \bibinfo {author} {\bibfnamefont {G.}~\bibnamefont
            {Gour}},\ }\bibfield  {title} {\bibinfo {title} {Quantum resource theories},\
    }\href {https://doi.org/10.1103/revmodphys.91.025001} {\bibfield  {journal}
        {\bibinfo  {journal} {Rev. Mod. Phys.}\ }\textbf {\bibinfo {volume} {91}},\
        \bibinfo {pages} {025001} (\bibinfo {year} {2019})}\BibitemShut {NoStop}%
    \bibitem [{\citenamefont {Helstrom}(1976)}]{Hel76}%
    \BibitemOpen
    \bibfield  {author} {\bibinfo {author} {\bibfnamefont {C.~W.}\ \bibnamefont
            {Helstrom}},\ }\href@noop {} {\emph {\bibinfo {title} {Quantum Detection and
                Estimation Theory}}}\ (\bibinfo  {publisher} {Academic},\ \bibinfo {address}
    {New York},\ \bibinfo {year} {1976})\BibitemShut {NoStop}%
    \bibitem [{\citenamefont {Holevo}(2011)}]{Hol11}%
    \BibitemOpen
    \bibfield  {author} {\bibinfo {author} {\bibfnamefont {A.}~\bibnamefont
            {Holevo}},\ }\href@noop {} {\emph {\bibinfo {title} {Probabilistic and
                Statistical Aspects of Quantum Theory}}}\ (\bibinfo  {publisher} {Edizioni
        della Normale},\ \bibinfo {address} {Pisa, Italy},\ \bibinfo {year}
    {2011})\BibitemShut {NoStop}%
    \bibitem [{\citenamefont {Zhang}\ \emph {et~al.}(2016)\citenamefont {Zhang},
                \citenamefont {Huang},\ and\ \citenamefont {Tong}}]{Zhang2016PRA}%
    \BibitemOpen
    \bibfield  {author} {\bibinfo {author} {\bibfnamefont {D.-J.}\ \bibnamefont
            {Zhang}}, \bibinfo {author} {\bibfnamefont {H.-L.}\ \bibnamefont {Huang}},\
        and\ \bibinfo {author} {\bibfnamefont {D.~M.}\ \bibnamefont {Tong}},\
    }\bibfield  {title} {\bibinfo {title} {{Non-Markovian quantum dissipative
                    processes with the same positive features as Markovian dissipative
                    processes}},\ }\href {https://doi.org/10.1103/PhysRevA.93.012117} {\bibfield
        {journal} {\bibinfo  {journal} {Phys. Rev. A}\ }\textbf {\bibinfo {volume}
            {93}},\ \bibinfo {pages} {012117} (\bibinfo {year} {2016})}\BibitemShut
    {NoStop}%
    \bibitem [{\citenamefont {Horn}\ and\ \citenamefont {Johnson}(2012)}]{HJ12}%
    \BibitemOpen
    \bibfield  {author} {\bibinfo {author} {\bibfnamefont {R.~A.}\ \bibnamefont
            {Horn}}\ and\ \bibinfo {author} {\bibfnamefont {C.~R.}\ \bibnamefont
            {Johnson}},\ }\href@noop {} {\emph {\bibinfo {title} {Matrix analysis}}}\
    (\bibinfo  {publisher} {Cambridge university press},\ \bibinfo {address}
    {Cambridge},\ \bibinfo {year} {2012})\BibitemShut {NoStop}%
    \bibitem [{\citenamefont {Paris}(2009)}]{PAR09}%
    \BibitemOpen
    \bibfield  {author} {\bibinfo {author} {\bibfnamefont {M.~G.~A.}\
            \bibnamefont {Paris}},\ }\bibfield  {title} {\bibinfo {title} {Quantum
            estimation for quantum technology},\ }\href
    {https://doi.org/10.1142/s0219749909004839} {\bibfield  {journal} {\bibinfo
            {journal} {Int. J. Quantum Inf.}\ }\textbf {\bibinfo {volume} {07}},\
        \bibinfo {pages} {125} (\bibinfo {year} {2009})}\BibitemShut {NoStop}%
    \bibitem [{\citenamefont {Proietti}\ \emph {et~al.}(2019)\citenamefont
                {Proietti}, \citenamefont {Ringbauer}, \citenamefont {Graffitti},
                \citenamefont {Barrow}, \citenamefont {Pickston}, \citenamefont {Kundys},
                \citenamefont {Cavalcanti}, \citenamefont {Aolita}, \citenamefont {Chaves},\
                and\ \citenamefont {Fedrizzi}}]{PRG19}%
    \BibitemOpen
    \bibfield  {author} {\bibinfo {author} {\bibfnamefont {M.}~\bibnamefont
            {Proietti}}, \bibinfo {author} {\bibfnamefont {M.}~\bibnamefont {Ringbauer}},
        \bibinfo {author} {\bibfnamefont {F.}~\bibnamefont {Graffitti}}, \bibinfo
        {author} {\bibfnamefont {P.}~\bibnamefont {Barrow}}, \bibinfo {author}
        {\bibfnamefont {A.}~\bibnamefont {Pickston}}, \bibinfo {author}
        {\bibfnamefont {D.}~\bibnamefont {Kundys}}, \bibinfo {author} {\bibfnamefont
            {D.}~\bibnamefont {Cavalcanti}}, \bibinfo {author} {\bibfnamefont
            {L.}~\bibnamefont {Aolita}}, \bibinfo {author} {\bibfnamefont
            {R.}~\bibnamefont {Chaves}},\ and\ \bibinfo {author} {\bibfnamefont
            {A.}~\bibnamefont {Fedrizzi}},\ }\bibfield  {title} {\bibinfo {title}
        {{Enhanced Multiqubit Phase Estimation in Noisy Environments by Local
                    Encoding}},\ }\href {https://doi.org/10.1103/physrevlett.123.180503}
    {\bibfield  {journal} {\bibinfo  {journal} {Phys. Rev. Lett.}\ }\textbf
        {\bibinfo {volume} {123}},\ \bibinfo {pages} {180503} (\bibinfo {year}
        {2019})}\BibitemShut {NoStop}%
    \bibitem [{\citenamefont {Zhang}\ \emph {et~al.}(2019)\citenamefont {Zhang},
                \citenamefont {Bromley}, \citenamefont {Huang}, \citenamefont {Cao},
                \citenamefont {Lv}, \citenamefont {Liu}, \citenamefont {Li}, \citenamefont
                {Guo}, \citenamefont {Cianciaruso},\ and\ \citenamefont {Adesso}}]{ZBH19}%
    \BibitemOpen
    \bibfield  {author} {\bibinfo {author} {\bibfnamefont {C.}~\bibnamefont
            {Zhang}}, \bibinfo {author} {\bibfnamefont {T.~R.}\ \bibnamefont {Bromley}},
        \bibinfo {author} {\bibfnamefont {Y.-F.}\ \bibnamefont {Huang}}, \bibinfo
        {author} {\bibfnamefont {H.}~\bibnamefont {Cao}}, \bibinfo {author}
        {\bibfnamefont {W.-M.}\ \bibnamefont {Lv}}, \bibinfo {author} {\bibfnamefont
            {B.-H.}\ \bibnamefont {Liu}}, \bibinfo {author} {\bibfnamefont {C.-F.}\
            \bibnamefont {Li}}, \bibinfo {author} {\bibfnamefont {G.-C.}\ \bibnamefont
            {Guo}}, \bibinfo {author} {\bibfnamefont {M.}~\bibnamefont {Cianciaruso}},\
        and\ \bibinfo {author} {\bibfnamefont {G.}~\bibnamefont {Adesso}},\
    }\bibfield  {title} {\bibinfo {title} {{Demonstrating Quantum Coherence and
                    Metrology that is Resilient to Transversal Noise}},\ }\href
    {https://doi.org/10.1103/physrevlett.123.180504} {\bibfield  {journal}
        {\bibinfo  {journal} {Phys. Rev. Lett.}\ }\textbf {\bibinfo {volume} {123}},\
        \bibinfo {pages} {180504} (\bibinfo {year} {2019})}\BibitemShut {NoStop}%
    \bibitem [{\citenamefont {Zhu}\ \emph {et~al.}(2022)\citenamefont {Zhu},
                \citenamefont {Zhang}, \citenamefont {Wang}, \citenamefont {Xiao},\ and\
                \citenamefont {Xue}}]{ZZW22}%
    \BibitemOpen
    \bibfield  {author} {\bibinfo {author} {\bibfnamefont {G.}~\bibnamefont
            {Zhu}}, \bibinfo {author} {\bibfnamefont {C.}~\bibnamefont {Zhang}}, \bibinfo
        {author} {\bibfnamefont {K.}~\bibnamefont {Wang}}, \bibinfo {author}
        {\bibfnamefont {L.}~\bibnamefont {Xiao}},\ and\ \bibinfo {author}
        {\bibfnamefont {P.}~\bibnamefont {Xue}},\ }\bibfield  {title} {\bibinfo
        {title} {Experimental witnessing for entangled states with limited local
            measurements},\ }\href {https://doi.org/10.1364/prj.462212} {\bibfield
        {journal} {\bibinfo  {journal} {Photon. Res.}\ }\textbf {\bibinfo {volume}
            {10}},\ \bibinfo {pages} {2047} (\bibinfo {year} {2022})}\BibitemShut
    {NoStop}%
    \bibitem [{\citenamefont {Watanabe}\ \emph {et~al.}(2010)\citenamefont
                {Watanabe}, \citenamefont {Sagawa},\ and\ \citenamefont {Ueda}}]{WSU10}%
    \BibitemOpen
    \bibfield  {author} {\bibinfo {author} {\bibfnamefont {Y.}~\bibnamefont
            {Watanabe}}, \bibinfo {author} {\bibfnamefont {T.}~\bibnamefont {Sagawa}},\
        and\ \bibinfo {author} {\bibfnamefont {M.}~\bibnamefont {Ueda}},\ }\bibfield
    {title} {\bibinfo {title} {{Optimal Measurement on Noisy Quantum Systems}},\
    }\href {https://doi.org/10.1103/physrevlett.104.020401} {\bibfield  {journal}
        {\bibinfo  {journal} {Phys. Rev. Lett.}\ }\textbf {\bibinfo {volume} {104}},\
        \bibinfo {pages} {020401} (\bibinfo {year} {2010})}\BibitemShut {NoStop}%
    \bibitem [{\citenamefont {Personick}(1971)}]{Per71}%
    \BibitemOpen
    \bibfield  {author} {\bibinfo {author} {\bibfnamefont {S.}~\bibnamefont
            {Personick}},\ }\bibfield  {title} {\bibinfo {title} {{Application of quantum
                    estimation theory to analog communication over quantum channels}},\ }\href
    {https://doi.org/10.1109/tit.1971.1054643} {\bibfield  {journal} {\bibinfo
            {journal} {IEEE Trans. Inf. Theory}\ }\textbf {\bibinfo {volume} {17}},\
        \bibinfo {pages} {240} (\bibinfo {year} {1971})}\BibitemShut {NoStop}%
    \bibitem [{\citenamefont {Gill}\ and\ \citenamefont {Levit}(1995)}]{GL95}%
    \BibitemOpen
    \bibfield  {author} {\bibinfo {author} {\bibfnamefont {R.~D.}\ \bibnamefont
            {Gill}}\ and\ \bibinfo {author} {\bibfnamefont {B.~Y.}\ \bibnamefont
            {Levit}},\ }\bibfield  {title} {\bibinfo {title} {{Applications of the van
                    Trees inequality: a Bayesian Cram{\'e}r-Rao bound}},\ }\href
    {https://doi.org/10.2307/3318681} {\bibfield  {journal} {\bibinfo  {journal}
            {Bernoulli}\ }\textbf {\bibinfo {volume} {1}},\ \bibinfo {pages} {59}
        (\bibinfo {year} {1995})}\BibitemShut {NoStop}%
    \bibitem [{\citenamefont {Hayashi}(2011)}]{Hay11}%
    \BibitemOpen
    \bibfield  {author} {\bibinfo {author} {\bibfnamefont {M.}~\bibnamefont
            {Hayashi}},\ }\bibfield  {title} {\bibinfo {title} {{Comparison Between the
                    Cramer-Rao and the Mini-max Approaches in Quantum Channel Estimation}},\
    }\href {https://doi.org/10.1007/s00220-011-1239-4} {\bibfield  {journal}
        {\bibinfo  {journal} {Commun. Math. Phys.}\ }\textbf {\bibinfo {volume}
            {304}},\ \bibinfo {pages} {689} (\bibinfo {year} {2011})}\BibitemShut
    {NoStop}%
    \bibitem [{\citenamefont {Macieszczak}\ \emph {et~al.}(2014)\citenamefont
    {Macieszczak}, \citenamefont {Fraas},\ and\ \citenamefont
    {Demkowicz-Dobrza{\'n}ski}}]{MFD14}%
    \BibitemOpen
    \bibfield  {author} {\bibinfo {author} {\bibfnamefont {K.}~\bibnamefont
        {Macieszczak}}, \bibinfo {author} {\bibfnamefont {M.}~\bibnamefont {Fraas}},\
    and\ \bibinfo {author} {\bibfnamefont {R.}~\bibnamefont
    {Demkowicz-Dobrza{\'n}ski}},\ }\bibfield  {title} {\bibinfo {title} {Bayesian
            quantum frequency estimation in presence of collective dephasing},\ }\href
    {https://doi.org/10.1088/1367-2630/16/11/113002} {\bibfield  {journal}
        {\bibinfo  {journal} {New J. Phys.}\ }\textbf {\bibinfo {volume} {16}},\
        \bibinfo {pages} {113002} (\bibinfo {year} {2014})}\BibitemShut {NoStop}%
    \bibitem [{\citenamefont {Rubio}\ and\ \citenamefont
                {Dunningham}(2019)}]{RD19}%
    \BibitemOpen
    \bibfield  {author} {\bibinfo {author} {\bibfnamefont {J.}~\bibnamefont
            {Rubio}}\ and\ \bibinfo {author} {\bibfnamefont {J.}~\bibnamefont
            {Dunningham}},\ }\bibfield  {title} {\bibinfo {title} {Quantum metrology in
            the presence of limited data},\ }\href
    {https://doi.org/10.1088/1367-2630/ab098b} {\bibfield  {journal} {\bibinfo
            {journal} {New J. Phys.}\ }\textbf {\bibinfo {volume} {21}},\ \bibinfo
        {pages} {043037} (\bibinfo {year} {2019})}\BibitemShut {NoStop}%
    \bibitem [{\citenamefont {Demkowicz-Dobrza{\'n}ski}\ \emph
    {et~al.}(2020)\citenamefont {Demkowicz-Dobrza{\'n}ski}, \citenamefont
    {G{\'o}recki},\ and\ \citenamefont {Gu{\c{t}}{\u{a}}}}]{DGG20}%
    \BibitemOpen
    \bibfield  {author} {\bibinfo {author} {\bibfnamefont {R.}~\bibnamefont
    {Demkowicz-Dobrza{\'n}ski}}, \bibinfo {author} {\bibfnamefont
        {W.}~\bibnamefont {G{\'o}recki}},\ and\ \bibinfo {author} {\bibfnamefont
        {M.}~\bibnamefont {Gu{\c{t}}{\u{a}}}},\ }\bibfield  {title} {\bibinfo {title}
        {{Multi-parameter estimation beyond quantum Fisher information}},\ }\href
    {https://doi.org/10.1088/1751-8121/ab8ef3} {\bibfield  {journal} {\bibinfo
            {journal} {J. Phys. A: Math. Theor.}\ }\textbf {\bibinfo {volume} {53}},\
        \bibinfo {pages} {363001} (\bibinfo {year} {2020})}\BibitemShut {NoStop}%
    \bibitem [{\citenamefont {G{\'o}recki}\ \emph {et~al.}(2020)\citenamefont
    {G{\'o}recki}, \citenamefont {Demkowicz-Dobrza{\'n}ski}, \citenamefont
    {Wiseman},\ and\ \citenamefont {Berry}}]{GDWB20}%
    \BibitemOpen
    \bibfield  {author} {\bibinfo {author} {\bibfnamefont {W.}~\bibnamefont
        {G{\'o}recki}}, \bibinfo {author} {\bibfnamefont {R.}~\bibnamefont
    {Demkowicz-Dobrza{\'n}ski}}, \bibinfo {author} {\bibfnamefont {H.~M.}\
        \bibnamefont {Wiseman}},\ and\ \bibinfo {author} {\bibfnamefont {D.~W.}\
        \bibnamefont {Berry}},\ }\bibfield  {title} {\bibinfo {title}
        {{{$\pi$}-Corrected Heisenberg Limit}},\ }\href
    {https://doi.org/10.1103/physrevlett.124.030501} {\bibfield  {journal}
        {\bibinfo  {journal} {Phys. Rev. Lett.}\ }\textbf {\bibinfo {volume} {124}},\
        \bibinfo {pages} {030501} (\bibinfo {year} {2020})}\BibitemShut {NoStop}%
    \bibitem [{\citenamefont {Rubio}\ and\ \citenamefont
                {Dunningham}(2020)}]{RD20}%
    \BibitemOpen
    \bibfield  {author} {\bibinfo {author} {\bibfnamefont {J.}~\bibnamefont
            {Rubio}}\ and\ \bibinfo {author} {\bibfnamefont {J.}~\bibnamefont
            {Dunningham}},\ }\bibfield  {title} {\bibinfo {title} {Bayesian
            multiparameter quantum metrology with limited data},\ }\href
    {https://doi.org/10.1103/physreva.101.032114} {\bibfield  {journal} {\bibinfo
            {journal} {Phys. Rev. A}\ }\textbf {\bibinfo {volume} {101}},\ \bibinfo
        {pages} {032114} (\bibinfo {year} {2020})}\BibitemShut {NoStop}%
    \bibitem [{\citenamefont {Sidhu}\ and\ \citenamefont {Kok}(2020)}]{SK20}%
    \BibitemOpen
    \bibfield  {author} {\bibinfo {author} {\bibfnamefont {J.~S.}\ \bibnamefont
            {Sidhu}}\ and\ \bibinfo {author} {\bibfnamefont {P.}~\bibnamefont {Kok}},\
    }\bibfield  {title} {\bibinfo {title} {Geometric perspective on quantum
            parameter estimation},\ }\href {https://doi.org/10.1116/1.5119961} {\bibfield
        {journal} {\bibinfo  {journal} {AVS Quantum Sci.}\ }\textbf {\bibinfo
            {volume} {2}},\ \bibinfo {pages} {014701} (\bibinfo {year}
        {2020})}\BibitemShut {NoStop}%
    \bibitem [{\citenamefont {G{\'o}recki}\ and\ \citenamefont
    {Demkowicz-Dobrza{\'n}ski}(2022)}]{GD22}%
    \BibitemOpen
    \bibfield  {author} {\bibinfo {author} {\bibfnamefont {W.}~\bibnamefont
        {G{\'o}recki}}\ and\ \bibinfo {author} {\bibfnamefont {R.}~\bibnamefont
    {Demkowicz-Dobrza{\'n}ski}},\ }\bibfield  {title} {\bibinfo {title}
        {{Multiple-Phase Quantum Interferometry: Real and Apparent Gains of Measuring
                    All the Phases Simultaneously}},\ }\href
    {https://doi.org/10.1103/physrevlett.128.040504} {\bibfield  {journal}
        {\bibinfo  {journal} {Phys. Rev. Lett.}\ }\textbf {\bibinfo {volume} {128}},\
        \bibinfo {pages} {040504} (\bibinfo {year} {2022})}\BibitemShut {NoStop}%
    \bibitem [{\citenamefont {Bavaresco}\ \emph {et~al.}(2024)\citenamefont
                {Bavaresco}, \citenamefont {Lipka-Bartosik}, \citenamefont {Sekatski},\ and\
                \citenamefont {Mehboudi}}]{BLSM24}%
    \BibitemOpen
    \bibfield  {author} {\bibinfo {author} {\bibfnamefont {J.}~\bibnamefont
            {Bavaresco}}, \bibinfo {author} {\bibfnamefont {P.}~\bibnamefont
            {Lipka-Bartosik}}, \bibinfo {author} {\bibfnamefont {P.}~\bibnamefont
            {Sekatski}},\ and\ \bibinfo {author} {\bibfnamefont {M.}~\bibnamefont
            {Mehboudi}},\ }\bibfield  {title} {\bibinfo {title} {{Designing optimal
                    protocols in Bayesian quantum parameter estimation with higher-order
                    operations}},\ }\href {https://doi.org/10.1103/physrevresearch.6.023305}
    {\bibfield  {journal} {\bibinfo  {journal} {Phys. Rev. Research}\ }\textbf
        {\bibinfo {volume} {6}},\ \bibinfo {pages} {023305} (\bibinfo {year}
        {2024})}\BibitemShut {NoStop}%
    \bibitem [{\citenamefont {Rubio}(2024)}]{Rub24}%
    \BibitemOpen
    \bibfield  {author} {\bibinfo {author} {\bibfnamefont {J.}~\bibnamefont
            {Rubio}},\ }\bibfield  {title} {\bibinfo {title} {First-principles
            construction of symmetry-informed quantum metrologies},\ }\href
    {https://doi.org/10.1103/physreva.110.l030401} {\bibfield  {journal}
        {\bibinfo  {journal} {Phys. Rev. A}\ }\textbf {\bibinfo {volume} {110}},\
        \bibinfo {pages} {{L030401}} (\bibinfo {year} {2024})}\BibitemShut {NoStop}%
    \bibitem [{\citenamefont {Zhou}\ \emph {et~al.}(2024)\citenamefont {Zhou},
                \citenamefont {Qiu},\ and\ \citenamefont {Zhang}}]{ZQZ24}%
    \BibitemOpen
    \bibfield  {author} {\bibinfo {author} {\bibfnamefont {Z.-Y.}\ \bibnamefont
            {Zhou}}, \bibinfo {author} {\bibfnamefont {J.-T.}\ \bibnamefont {Qiu}},\ and\
        \bibinfo {author} {\bibfnamefont {D.-J.}\ \bibnamefont {Zhang}},\ }\bibfield
    {title} {\bibinfo {title} {{S}trict hierarchy of optimal strategies for
    global estimations: {L}inking global estimations with local ones},\ }\href
    {https://doi.org/10.1103/physrevresearch.6.l032048} {\bibfield  {journal}
        {\bibinfo  {journal} {Phys. Rev. Research}\ }\textbf {\bibinfo {volume}
            {6}},\ \bibinfo {pages} {{L032048}} (\bibinfo {year} {2024})}\BibitemShut
    {NoStop}%
    \bibitem [{\citenamefont {Simon}(1996)}]{Sim97}%
    \BibitemOpen
    \bibfield  {author} {\bibinfo {author} {\bibfnamefont {B.}~\bibnamefont
            {Simon}},\ }\href@noop {} {\emph {\bibinfo {title} {Representations of finite
                and compact groups}}},\ Vol.~\bibinfo {volume} {10}\ (\bibinfo  {publisher}
    {American Mathematical Society},\ \bibinfo {address} {Providence, Rhode
        Island, USA},\ \bibinfo {year} {1996})\BibitemShut {NoStop}%
    \bibitem [{\citenamefont {Kimura}(2003)}]{Kim03}%
    \BibitemOpen
    \bibfield  {author} {\bibinfo {author} {\bibfnamefont {G.}~\bibnamefont
            {Kimura}},\ }\bibfield  {title} {\bibinfo {title} {{The Bloch vector for
                        {$N$}-level systems}},\ }\href
    {https://doi.org/10.1016/s0375-9601(03)00941-1} {\bibfield  {journal}
        {\bibinfo  {journal} {Phys. Lett. A}\ }\textbf {\bibinfo {volume} {314}},\
        \bibinfo {pages} {339} (\bibinfo {year} {2003})}\BibitemShut {NoStop}%
    \bibitem [{\citenamefont {Watrous}(2018)}]{Wat18}%
    \BibitemOpen
    \bibfield  {author} {\bibinfo {author} {\bibfnamefont {J.}~\bibnamefont
            {Watrous}},\ }\href@noop {} {\emph {\bibinfo {title} {The Theory of Quantum
                Information}}}\ (\bibinfo  {publisher} {Cambridge University Press},\
    \bibinfo {year} {2018})\BibitemShut {NoStop}%
\end{thebibliography}
%

\end{document}